\documentclass[prd,showpacs,twocolumn,superscriptaddress,floatfix,10pt]{revtex4-2}
\usepackage{graphicx}
\usepackage{amsfonts,amssymb,stmaryrd,latexsym,amsmath}
\usepackage[colorlinks,citecolor=blue,urlcolor=blue,linkcolor=blue]{hyperref}
\usepackage{slashed}
\usepackage{orcidlink}
\usepackage{bm}
\usepackage{multirow}
\usepackage{bbm}
\usepackage[normalem]{ulem}
\usepackage{mdframed}

\allowdisplaybreaks

\newcommand{\Tr}{\text{Tr}}
\newcommand{\etal}{\textit{et al}. }

\def\DD{{DD^{*}}}
\def\DDbar{{D\bar{D}^{*}/\bar{D}D^*}}

\newcommand{\ttle}{Identification of the $G(3900)$ as the P-wave $D\bar{D}^*/\bar{D}D^*$  resonance}

\newcommand{\smabs}{the one-boson exchange (OBE) potentials, identifying $G(3900)$ pole in the $K$-matrix formalism, the systemic uncertainties and the production line shape of $D^*D$ states}

\begin{document}



\title{\ttle }

 \newcommand{\rub}{\affiliation{Institut f\"ur Theoretische Physik II, Ruhr-Universit\"at Bochum, D-44780 Bochum, Germany }}
\newcommand{\pkusp}{\affiliation{School of Physics, Peking University, Beijing 100871, China}}
\newcommand{\pku}{\affiliation{School of Physics and Center of High Energy Physics, Peking University, Beijing 100871, China}}

\author{Zi-Yang Lin\,\orcidlink{0000-0001-7887-9391}}\thanks{These authors equally contribute to this work.}\pkusp

\author{Jun-Zhang Wang\,\orcidlink{0000-0002-3404-8569}}\thanks{These authors equally contribute to this work.}\pku

\author{Jian-Bo Cheng\orcidlink{0000-0003-1202-4344}}%
\affiliation{College of Science, China University of Petroleum, Qingdao, Shandong 266580, China}
\author{Lu Meng\, \orcidlink{0000-0001-9791-7138}}\email{Corresponding author, Email: lu.meng@rub.de}\rub

\author{Shi-Lin Zhu\,\orcidlink{0000-0002-4055-6906}}\email{Corresponding author, Email: zhusl@pku.edu.cn}\pku

\begin{abstract}

The BESIII Collaboration recently performed a precise measurement of the $e^+e^-\rightarrow D\bar{D}$ Born cross sections, and confirmed the $G(3900)$ structure reported by BaBar and Belle with high significance. We identify the $G(3900)$ as the first P-wave $\DDbar$ molecular resonance. The experimental and theoretical identification of the P-wave dimeson state holds paramount importance in enhancing our comprehension of the non-perturbative QCD and few-body physics. Its existence is firmly established in a unified meson-exchange model which simultaneously depicts the features of the $\chi_{c1}(3872)$, $Z_c(3900)$ and $T_{cc}(3875)$. This scenario can be directly examined in the $e^+e^-\rightarrow D\bar{D}^*/\bar{D}D^*$ cross section by seeing whether a resonance exists at the threshold. The credibility of the investigations is also ensured by the fact that the P-wave interaction dominantly arises from the well-known long-range pion exchange. Additionally, thanks to the centrifugal barrier, it is easier to form resonances in P-wave than in S-wave. We extensively calculate all systems up to P-wave with various quantum numbers and predict a dense population of the $\DDbar$ and $\DD$ states, where the S-wave $\DDbar$ state with $I^G (J^{PC})=0^- (1^{+-})$, P-wave $\DDbar$ state with $I^G(J^{PC})=0^+(0^{-+})$, and P-wave $\DD$ state with $I(J^P)=0(0^-)$ are more likely to be observed in experiments. 

\end{abstract}

\maketitle

\emph{Introduction.}---   Over the past two decades, a significant number of  hadrons defying the spectra predicted by quark models have been observed in the heavy flavor sector,  which are typically regarded as the exotica within the realm of Quantum Chromodynamics (QCD), see Refs.~\cite{Chen:2016qju,Esposito:2016noz,Guo:2017jvc,Liu:2019zoy,Brambilla:2019esw,Chen:2022asf,Meng:2022ozq} for reviews.  Delving into the structure and dynamics associated with these exotic states holds paramount importance in enhancing our comprehension of the non-perturbative features of low-energy QCD. These states also serve as promising examples for studying the general few-body physics. 

Among these exotic states, the $\chi_{c1}(3872)$, $Z_c(3900)$, and $T_{cc}(3875)$ stand out as undeniable ``star" examples, believed to be the first charmonium-like state~\cite{Belle:2003nnu}, the first manifestly exotic charmonium-like state~\cite{BESIII:2013ris,Belle:2013yex}, and the first doubly charmed tetraquark state~\cite{LHCb:2021vvq,LHCb:2021auc} observed in experiments, respectively. It is particularly intriguing that these three states are closely interconnected. The proximity of the former two states to the $\DDbar$ threshold and the latter one to the $\DD$ threshold positions them as strong candidates for corresponding hadronic molecules. Indeed, prior to the observation of $T_{cc}(3875)$, Li \etal  had predicted a very loosely bound state of $\DD$ utilizing the one-boson-exchange model (OBE), with parameters established beforehand while investigating the $\chi_{c1}(3872)$~\cite{Li:2012cs,Li:2012ss}. It is noteworthy that although the doubly heavy tetraquark states have been anticipated for over 40 years~\cite{Ader:1981db} (see a brief review in Ref.~\cite{Richard:2022fdc}), there was little literature explicitly pointing out the existence of molecular-type doubly charmed tetraquark states before the LHCb results~\cite{LHCb:2021vvq,LHCb:2021auc}.

In the realm of doubly heavy exotic states, such as the $\chi_{c1}(3872)$, $Z_c(3900)$, and $T_{cc}(3875)$, previous studies have predominantly focused on S-wave dimeson states, encompassing bound states, virtual states, or resonances. However, P-wave states near the threshold are of particular interest and arouse the attention in many fields of physics, see the halo nuclei as P-wave resonances in nuclear physics~\cite{Bertulani:2002sz} and the P-wave Feshbach resonances in cold atomic physics~\cite{chin2010feshbach}. In the context of effective field theory, P-wave and higher partial wave systems were employed to challenge the conventional Weinberg power counting rule~\cite{Nogga:2005hy}, sparking intense debate on the renormalization of non-perturbative effective field theory, (see Ref.~\cite{Epelbaum:2009sd} and references therein). Recently, the BESIII Collaboration performed a precise measurement of Born cross sections for the $e^+e^-\rightarrow D\bar{D}$ process ~\cite{BESIII:2024ths}. Apart from the established $1^{--}$ states $\psi(3770)$, $\psi(4040)$, $\psi(4160)$, $\psi(4230)$, $\psi(4360)$, $\psi(4415)$, $\psi(4660)$, they observed a peak structure around 3.9 GeV. Its mass and width are fitted in the Breit-Wigner formalism to be $3872.5\pm 14.2\pm 3.0$ MeV and $179.7\pm 14.1\pm 7.0$ MeV, respectively (see the Supplemental Materials of Ref.~\cite{BESIII:2024ths}). 
It is worth noting that in previous works~\cite{BaBar:2006qlj,BaBar:2008drv,Belle:2007qxm}, this peak was not treated as a resonance, and thus named $G(3900)$.
The coupled-channel analysis of data from Belle and BESIII has the potential to generate a bump at this position without introducing new states. However, it appears to be very challenging in accurately depicting the nearby points ~\cite{Uglov:2016orr,Nakamura:2023obk}. In this letter, for simplicity, we will consistently refer to this structure as $G(3900)$, even if it corresponds to a genuine resonance state.

Very recently, a $K$-matrix formalism has been used to analyze the new $e^+e^-\to D\bar{D}$ data from BESIII, along with other inclusive and exclusive data, concluding that no additional bare pole near 3900 MeV is needed~\cite{Husken:2024hmi}. However, the $K$-matrix formalism adopted in Ref.~\cite{Husken:2024hmi} differs from the standard one presented in the Review of Particle Physics~\cite{ParticleDataGroup:2024cfk}. We identified that the amplitude formalism in Ref.~\cite{Husken:2024hmi} does not fulfill the requirement of analyticity, causing problems when continuing the $K$-matrix into the complex plane. After correcting these defects and refitting the same data using a $K$-matrix formalism that fulfills analyticity, unitarity, and threshold behaviors, we find robust evidence for the existence of the $G(3900)$ pole~\cite{Suppl}. Du \etal analyzed $e^+e^-\to D\bar{D}$ data from Belle~\cite{Belle:2007qxm} using the Lippmann-Schwinger equation formalism and obtained the pole $3.879-i0.035$ GeV corresponding to the $G(3900)$~\cite{Du:2016qcr}, which agrees well with our fit result.

In this work, we aim to identify the $G(3900)$ as the first P-wave dimeson state in the doubly heavy sector in the meson-exchange model. By relating the $G(3900)$ to the S-wave states $\chi_{c1}(3872)$, $Z_c(3900)$ and $T_{cc}^+(3985)$, we make a unified description of these states in the one-boson-exchange (OBE) interaction. The resonance poles are obtained by solving the complex scaled Schr\"odinger equation in momentum space (The details can be found in Ref.~\cite{Lin:2023ihj}). For the following three reasons, the predictions regarding the existence of the P-wave resonance are highly reliable.

Just as the OBE model has provided a high-precision description of nuclear forces~\cite{Machleidt:1987hj}, meson-exchange models have also achieved notable success in elucidating heavy flavor hadronic molecules~\cite{Tornqvist:1991ks,Tornqvist:1993ng,Liu:2009qhy,Liu:2008fh,Liu:2008xz,Ding:2009vj,Sun:2011uh,Thomas:2008ja,Lee:2009hy}. In the 1990s, Törnqvist predicted a deuteron-like $D\bar{D}^*$ bound state, which has been confirmed by the observation of the $\chi_{c1}(3872)$~\cite{Tornqvist:1991ks,Tornqvist:1993ng}. The interactions stemming from the exchange of $\pi$, $\eta$, $\rho$, $\omega$, and $\sigma$ particles naturally predict $T_{cc}$ as a $\DD$ bound state once their parameters are determined in the $\DDbar$ systems, specifically the $\chi_{c1}(3872)$~\cite{Li:2012cs,Li:2012ss}. The interactions governing $T_{cc}$ and $\chi_{c1}(3872)$ adhere to the G-parity rules. Given that both P-wave and S-wave states arise from the partial wave expansion of the same potential, the existence of these P-wave resonances could be firmly established once the S-wave interaction is fixed in depicting the $\chi_{c1}(3872)$, $Z_c(3900)$ and $T_{cc}^+(3985)$ states.  Recently, similar ideas have been used to investigate the P-wave $D\bar{D}_1$, $D^*\bar{D}_1$ and $D^*\bar{D}_2^*$ states~\cite{Wang:2023ivd}, inspired by the corresponding deeply bound S-wave states in Ref.~\cite{Ji:2022blw}. The states of $D^*N$ in the P-wave were also considered during the investigation of $\Lambda_c(2940)^+$ and its counterparts~\cite{He:2010zq}.

\begin{figure}
    \centering
    \includegraphics[width=0.45\textwidth]{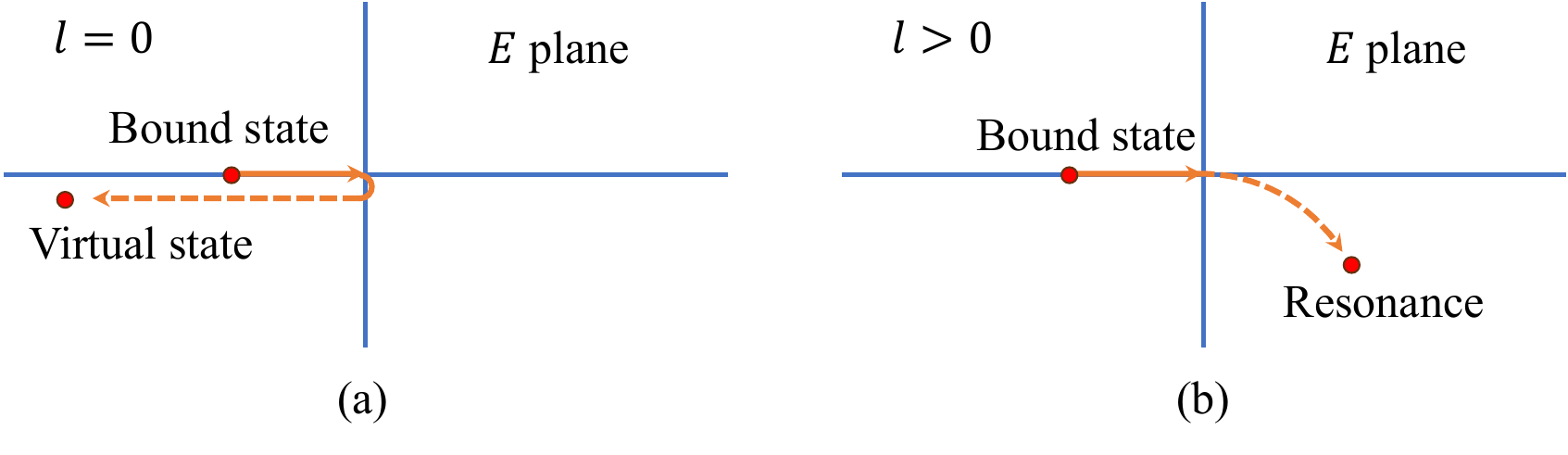}
    \caption{Transition of the bound state pole to the virtual state (a) and resonance (b) for S-wave and higher partial waves, respectively by adjusting the strength of the potential to be less attractive~\cite{Taylor:1972pty}. The solid (dashed) lines represent the pole trajectories in the physical (unphysical) Riemann sheets.}
    \label{fig:pole_taylor}
\end{figure}

Basically, the generation of S-wave shape-type resonances typically hinges on the specific dynamical mechanisms, for example an attractive potential with a repulsive barrier. Conversely, P-wave resonances can be generated more easily thanks to the centrifugal barrier.  As shown in  Fig.~\ref{fig:pole_taylor}, in a single-channel system, as the potential becomes less attractive, the P-wave (or higher partial waves) bound state pole in the physical Riemann sheet tends to migrate into the unphysical sheet manifesting as a resonance \cite{Taylor:1972pty}, while the S-wave bound state becomes a virtual state. In the P-wave case, the centrifugal barrier provides a repulsion and naturally leads to the formation of resonances.

The dynamics of the P-wave system is more dependent on the long-range interaction than the S-wave ones. Given the repulsion effect of the centrifugal barrier, the particles tend to keep away from the origin, rendering the short-range interactions become less important, and the peripheral interactions, specifically the  well-known one-pion-exchange (OPE) potential, will play a vital role. Indeed, taking the chiral effective field theory as an example, the leading-order chiral interaction of the P-wave system solely stems from the OPE interaction, while for the S-wave systems, the OPE interaction is accompanied with contact interactions to ensure the renormalization~\cite{Epelbaum:2008ga,Baru:2015nea,Du:2021zzh,Wang:2022jop,Meng:2019ilv,Lin:2022wmj,Meng:2022ozq}. Similar conclusions were also supported by analyzing the lattice data~\cite{Meng:2023bmz}. In the literature, there are some variants of the OBE model~\cite{Voloshin:1976ap,Tornqvist:1991ks,Tornqvist:1993ng} on the short-range interactions, while the coupling constants of the long-range OPE interaction have been determined by the partial decay width of $D^*$. Consequently, P-wave resonances are typically predicted with high credibility.

\emph{Framework.}---  We adopt a framework established in Refs.~\cite{Georgi:1990cx,Mannel:1990vg,Falk:1991nq,Liu:2008fh,Liu:2009qhy,Li:2012cs,Li:2012ss}. Under the heavy quark spin symmetry, the pseudoscalar $D$, the vector meson $D^*$ and their antiparticles are combined into the superfield $\mathcal{H}$ and $\tilde{\mathcal{H}}$,

\begin{eqnarray}
    {\cal H}=\frac{1+\slashed{v}}{2}(P_{\mu}^*\gamma^{\mu}-P\gamma_{5}),\,\tilde{{\cal H}}=(\tilde{P}_{\mu}^*\gamma^{\mu}-\tilde{P}\gamma_{5})\frac{1-\slashed{v}}{2},
\end{eqnarray}
where $P=(D^0,D^+)$, $P_\mu^*=(D^{*0},D^{*+})_\mu$, $\tilde{P}=(\bar{D}^0,D^-)^\text{T}$, $\tilde{P}_\mu^*=(\bar{D}^{*0},D^{*-})_\mu^\text{T}$. $v=(1,0,0,0)$ is the velocity of the heavy meson. The conjugation of $\cal{H}$ and $\tilde{\mathcal{H}}$ is defined as $\bar{\mathcal{H}}=\gamma_0\mathcal{H}^\dagger\gamma_0$ and $\bar{\tilde{\mathcal{H}}}=\gamma_0\tilde{\mathcal{H}}^\dagger\gamma_0$. For the charge conjugation transformation, we adopt the convention $D\xrightarrow{\mathcal{C}}\bar{D}$ and $D^*\xrightarrow{\mathcal{C}}-\bar{D}^*$, namely $\mathcal{H}\xrightarrow{\mathcal{C}}C^{-1}\tilde{\mathcal{H}}^\text{T}C$, where $C=i\gamma^2\gamma^0$. We include the  $\pi$, $\eta$, $\sigma$, $\rho$, $\omega$ exchanges in the OBE model via the following Lagrangians,

\begin{eqnarray}
    &\mathcal{L}&=g_s\Tr\left[\mathcal{H}\sigma\bar{\mathcal{H}}\right]+ig_a\Tr\left[\mathcal{H}\gamma_\mu\gamma_5\mathcal{A}^\mu\bar{\mathcal{H}}\right]\nonumber\\
    &&+i\beta\Tr\left[\mathcal{H} v_\mu (\mathcal{V}^\mu-\rho^\mu)\bar{\mathcal{H}}\right]+i\lambda\Tr\left[\mathcal{H}\sigma_{\mu\nu}F^{\mu\nu}\bar{\mathcal{H}}\right]\nonumber\\
    && + \text{c.c. terms } (\mathcal{H}\to\bar{\tilde{\mathcal{H}}},\bar{\mathcal{H}}\to \tilde{\mathcal{H}} ).
\end{eqnarray}
The vector meson fields $\rho^\mu$ and the pseudoscalar meson fields $\mathbb{P}$ are defined as
\begin{eqnarray}
 \rho^\mu=\frac{ig_V}{\sqrt{2}}\begin{bmatrix}\frac{\rho^{0}+\omega}{\sqrt{2}}
 & \rho^{+}\\
\rho^{-} & \frac{-\rho^{0}+\omega}{\sqrt{2}}
\end{bmatrix}^\mu,
\mathbb{P}=
\begin{bmatrix}\frac{\pi^{0}}{\sqrt{2}}+\frac{\eta}{\sqrt{6}} & \pi^{+}\\
\pi^{-} & -\frac{\pi^{0}}{\sqrt{2}}+\frac{\eta}{\sqrt{6}}
\end{bmatrix}.\nonumber\\
\end{eqnarray}
$F^{\mu\nu}=\partial^\mu\rho^\nu-\partial^\nu\rho^\mu-[\rho^\mu,\rho^\nu]$ represents the field strength tensor of vector mesons. $\mathcal{V}^\mu$ and $\mathcal{A}^\mu$ represent the vector and axial currents of pseudoscalar mesons, respectively
\begin{eqnarray}
    &&\mathcal{V}^\mu=\frac{1}{2}[\xi^\dagger,\partial_\mu\xi],\; \mathcal{A}^\mu=\frac{1}{2}\{\xi^\dagger,\partial_\mu\xi\},\;
    \xi=\exp(i\mathbb{P}/f_\pi).
\end{eqnarray} 
$f_\pi=132$ MeV is the pion decay constant.  The isospin average masses of particles are taken from the Review of Particle Physics \cite{ParticleDataGroup:2024cfk}: $m_\pi=137$ MeV, $m_\eta=548$ MeV, $m_\rho=775$ MeV, $m_\omega=783$ MeV, $m_D=1867$ MeV, $m_{D^*}=2009$ MeV. The axial coupling constant $g_a=0.59$ is extracted from the $D^*$ width. The coupling constants are fixed to be consistent with Refs.~\cite{Li:2012ss, Li:2012cs}, resulting in an accurate depiction of $\chi_{c1}(3872)$ and a remarkable prediction of the $T_{cc}$ state. The couplings in the vector-meson-exchange process are determined by vector meson dominance, combined with lattice QCD and light cone sum rules~\cite{Isola:2003fh}: $g_V = 5.8$, $\beta = 0.9$, and $\lambda = 0.56 \text{ GeV}^{-1}$. For the scalar meson exchange, the mass and coupling are derived from the $\Sigma$-model~\cite{Bardeen:2003kt, Liu:2008xz}: $m_\sigma = 600$ MeV and $g_s = 0.76$. To estimate the uncertainties of the coupling constants, we redetermine the three independent coupling constants by the pole positions of $\chi_{c1}(3872)$, $T_{cc}^+(3985)$, and $Z_c(3900)$ in the Supplemental Materials. The results show no qualitative differences~\cite{Suppl}.

\begin{figure}
    \centering
    \includegraphics[width=0.48\textwidth]{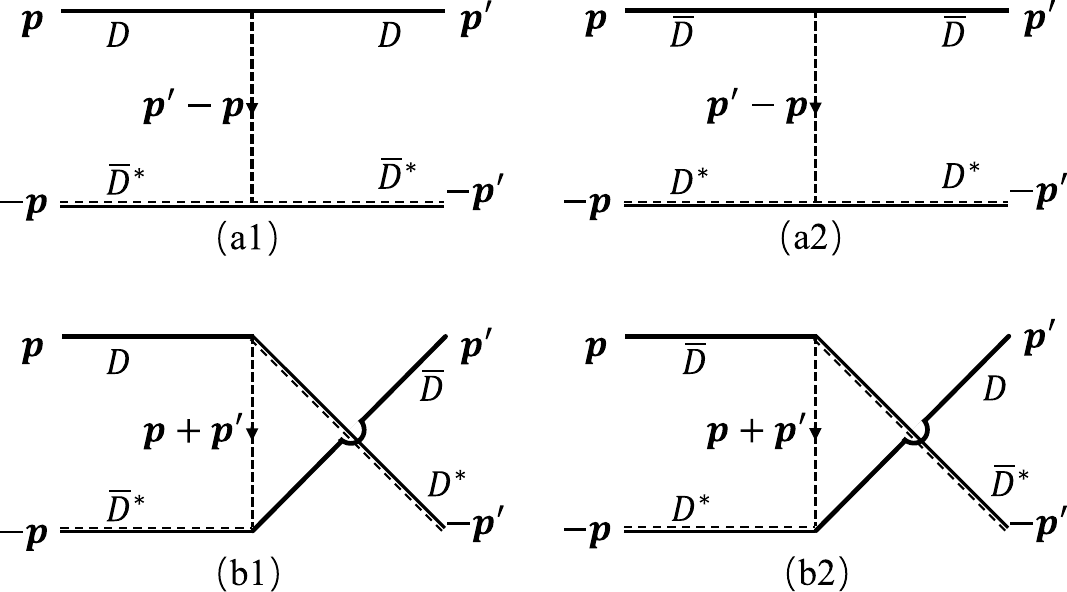} 
    \caption{\label{fig:OBE} The direct diagrams (upper row) and cross diagrams (lower row) in the OBE, where the transferred momenta are $\bm{p}'-\bm p$ and $\bm p'+\bm p$, respectively.}
\end{figure}

We construct the $D\bar{D}^*$ wave functions as the C-parity eigenstates for neutral channels,
\begin{eqnarray}
    |C=\pm\rangle=\frac{1}{\sqrt{2}}(|D(\bm{p})\bar{D}^*(-\bm{p})\rangle\mp|\bar{D}(\bm{p})D^*(-\bm{p})\rangle).
\end{eqnarray}
 For the charged channels, we can similarly construct the wave functions as the eigenstates of G-parity. As depicted in Fig.~\ref{fig:OBE}, the transfer momentum in the cross diagrams corresponding to the $u$-channel turns out $\bm{k}=\bm{p}+\bm{p}'$, while it is $\bm{q}=\bm{p}-\bm{p}'$ in the direct diagrams or $t$-channel exchanges. It is explained in 
Supplemental Material~\cite{Suppl}
that the momentum labeling is crucial to get the correct P-wave interactions. To show the key mechanism of the P-wave resonance, we ignore the isospin breaking effect and adopt the time-component of $q^0,k^0=0$. 

The effective potential for the $\DDbar$ system can be related to the $\DD$ potential up to a factor $G_{m}$ ($-G_m G_{MM}$) for the direct (cross) diagrams, with $G_m$ and $G_{MM}$ as the G-parities of the exchanged meson and the $\DDbar$ system, respectively. It is noticeable that the G-parity rule for cross diagrams is different from that for direct diagrams.  The specific effective potentials and the derivation of the G-parity rule are presented in 
Supplemental Material~\cite{Suppl}.
With the complex scaling method $p\rightarrow pe^{-i\theta}$, the resonance and bound state poles can be derived as the eigenenergy in the Schrödinger equation
\begin{eqnarray}
    E\phi(\bm{p})=\frac{\bm{p}^2}{2\mu}\phi(\bm{p})+\int V(\bm{p},\bm{{k}})\phi(\bm{k})\frac{d^3\bm{k}}{(2\pi)^3}.
\end{eqnarray}
To search for virtual states, we adopt the method in Ref.~\cite{Chen:2023eri}.

To regularize the ultraviolet divergence in the integral, we introduce a monopole regulator to suppress the potential at the large momentum
\begin{eqnarray}
    V(\bm{p}',\bm{p})\rightarrow V(\bm{p}',\bm{p})\frac{\Lambda^2}{p'^2+\Lambda^2}\frac{\Lambda^2}{p^2+\Lambda^2}.
\end{eqnarray}
The dependence on the regulator is investigated in Supplemental Materials, and our final results remain consistent regardless of the specific regulator chosen. The cutoff $\Lambda$ is the only parameter to be determined. We adjust $\Lambda$ to generate a pole at the threshold (a loosely bound state or a near-threshold virtual state) in the $^3S_1$ isosinglet $\DDbar$ system with the positive C-parity for neutral components, namely the $1^{++}$ channel corresponding to $\chi_{c1}(3872)$. Then we search for poles with different isospins, C-parities, orbital angular momenta (S-wave and P-wave) in $\DDbar$ and $\DD$ systems. 

   \begin{figure*}
       \centering
       \includegraphics[width=1\textwidth]{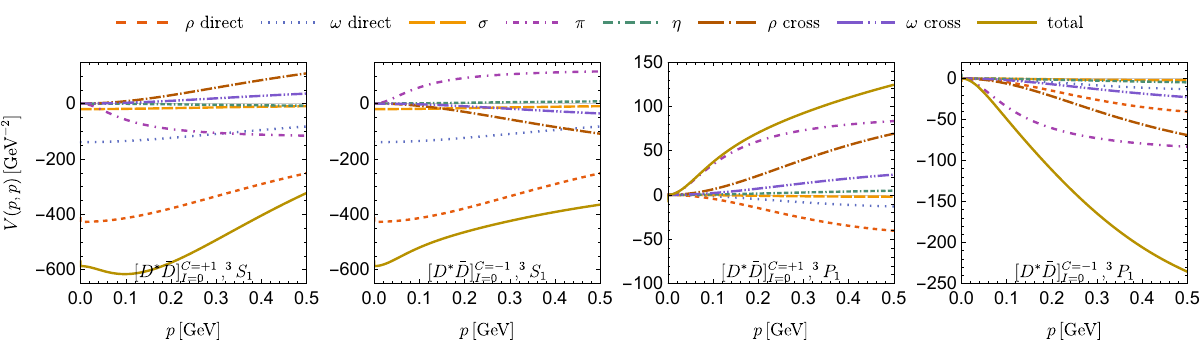}
       \caption{\label{fig:potential}The OBE potentials in the $1^{++}$, $1^{+-}$, $1^{-+}$ and $1^{--}$ isoscalar $\DDbar$ channels without regulators. The $1^{++}$ and $1^{-+}$ channel correspond to the $\chi_{c1}(3872)$ and its P-wave counterpart, respectively. The $1^{--}$ and $1^{+-}$ channel corresponds to the $G(3900)$ and its S-wave counterpart, respectively. Only $p=p'$ cases are shown. }
   \end{figure*}

\emph{Results and discussion.}--- 
The partial-wave potentials of $J=1$ isosinglet $\DDbar$ systems are depicted in Fig.~\ref{fig:potential}. In P-wave interactions, due to the centrifugal barrier, the significance of the long-range pion exchange increases, whereas the S-wave interaction is predominantly governed by the $\rho$ exchange. The $\chi_{c1}(3872)$ corresponds to the $1^{++}$ S-wave channel, exhibiting the most pronounced attraction. Its negative C-parity counterpart, the $1^{+-}$ S-wave channel, also displays an attractive potential. Consequently, these two S-wave channels may give rise to near-threshold bound states or virtual states. The $1^{-+}$ channel, serving as the P-wave counterpart of $1^{++}$, demonstrates substantial repulsion, thus making it unlikely to produce poles near the threshold. However, the potential of the $1^{--}$ channel, which is the P-wave partner of the $1^{+-}$ channel, is attractive, suggesting a possible resonance pole corresponding to the $G(3900)$.

\begin{figure}[htp]
    \includegraphics[width=0.48\textwidth]{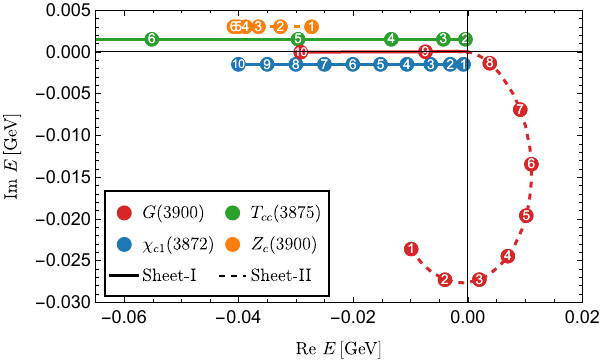}
    \caption{The pole trajectories with the cutoff parameters correspond to $\chi_{c1}(3872)$, $T_{cc}(3875)$, $Z_{c}(3900)$ and the newly observed $G(3900)$ states. The circled number 1-10 represent the increasing cutoff 0.4-1.3 GeV in order. The solid (dashed) lines represent the pole trajectories in the physical (unphysical) Riemann sheets. The poles on the negative real axis are slightly shifted  for transparency.}\label{fig:pole}
\end{figure}

In Fig.~\ref{fig:pole}, we illustrate the pole trajectories of four particularly intriguing states: the $\chi_{c1}(3872)$, $T_{cc}(3875)$, $Z_{c}(3900)$, and the recently observed $G(3900)$, as the cutoff parameter $\Lambda$ varies from 0.4 GeV to 1.3 GeV. With a cutoff of around 0.5 GeV, the $\chi_{c1}(3872)$ manifests as a loosely bound state. The $T_{cc}$ is also a near-threshold bound state, which agrees with the results in Ref.~\cite{Li:2012ss}. Simultaneously, the $Z_c(3900)$ emerges as a virtual state, aligning with the pole position deduced through a data-driven coupled-channel analysis in Ref.~\cite{Nakamura:2023obk}. Remarkably, within this same cutoff range, a P-wave resonance materializes in the $1^{--}$ channel, corresponding to the $G(3900)$ state. If the cutoff is increased to strengthen the attraction, the $G(3900)$ resonance will move to the physical Riemann sheet and turns into a bound state, thereby confirming the $G(3900)$ as indeed a P-wave resonance engendered by adjusting the interaction strength. The value of the cutoff differs from the results in Refs.~\cite{Li:2012ss,Cheng:2022qcm}, since the regulator is different. However, our conclusion holds under different regulators. We test the results using the regulator and cutoff fixed in Ref.~\cite{Cheng:2022qcm}. We validate these findings using the regulator and cutoff parameters established in Ref.~\cite{Cheng:2022qcm}. The results indicate that as long as the cutoff is set to generate a loosely bound $\chi_{c1}(3872)$ state, a corresponding P-wave resonance emerges in the $1^{--}$ channel as the $G(3900)$, while the poles of $T_{cc}$ and $Z_c(3900)$ remain qualitatively unchanged, see the 
Supplemental Materials~\cite{Suppl}.

\begin{table*}
    \centering
    \caption{\label{tab:prediction} The poles in all channels of $D\bar{D}^*$ and $DD^*$, up to the orbital angular momentum $L=1$ (in unit of MeV). The $B$ and $V$ superscripts denote the bound state and the virtual state, respectively. Otherwise the pole refers to a resonance. }
    \begin{tabular*}{\hsize}{@{}@{\extracolsep{\fill}}cccccccc@{}}
        \hline \hline 
         &  & \multicolumn{2}{c}{$D\bar{D}^{*}\,,C=+$} & \multicolumn{2}{c}{$D\bar{D}^{*}\,,C=-$} & \multicolumn{2}{c}{$DD^{*}$}\tabularnewline
         &  & $I=0$ & $I=1$ & $I=0$ & $I=1$ & $I=0$ & $I=1$\tabularnewline
        \hline 
        \multirow{4}{*}{$\Lambda=0.5$GeV} & $1^{+}({}^{3}S_{1})$ & $-3.1^B$, $\chi_{c1}(3872)$ & - & $-1.60^B$ & $-34.8^V$, $Z_{c}(3900)$ & $-0.41^B$, $T_{cc}(3875)$ & -\tabularnewline
         & $0^{-}({}^{3}P_{0})$ & $-1.5-14.5i$ & - & - & - & $-9.6-9.7i$ & -\tabularnewline
         & $1^{-}({}^{3}P_{1})$ & - & - & $-4.0-27.3i$, $G(3900)$ & - & $-31.7-70.6i$ & - \tabularnewline
         & $2^{-}({}^{3}P_{2})$ & $-42.6-39.4i$ & - & $-21.3-50.7i$ & - & $-37.8-40.9i$ & - \tabularnewline
        \hline 
        \multirow{4}{*}{$\Lambda=0.6$GeV} & $1^{+}({}^{3}S_{1})$ & $-6.5^B$, $\chi_{c1}(3872)$ & - & $-5.8^B$ & $-39.5^V$, $Z_{c}(3900)$ & 
                $-4.3^B$, $T_{cc}(3875)$ & -\tabularnewline
         & $0^{-}({}^{3}P_{0})$ & $3.2-13.7i$ & - & - & - & $-10.2-12.1i$ & -\tabularnewline
         & $1^{-}({}^{3}P_{1})$ & - & - & $2.0-27.3i$, $G(3900)$ & - & $-33.7-84.8i$ & -\tabularnewline
          & $2^{-}({}^{3}P_{2})$ & $-44.2-49.0i$ & - & $-19.3-58.8i$ & - & $-37.8-49.3i$ & -\tabularnewline
        \hline         \hline 
        \end{tabular*}
\end{table*}

We delve into the resonances with alternative quantum numbers, as summarized in Table~\ref{tab:prediction}. We neglect the tiny imaginary part of virtual state pole arsing from the left-hand cut of the pion exchange. For $\DDbar$ isospin singlets, aside from the $\chi_{c1}(3872)$ and $G(3900)$, a loosely bound state exists in the S-wave partner channel of the $G(3900)$. Although the real part of the P-wave $G(3900)$ pole position is close to that of its S-wave partner, they cannot be considered nearly degenerate due to the significant width of the $G(3900)$. In channels with $J=0$ and $J=2$, resonances emerge in $0^{-+}$, $2^{--}$, and $2^{-+}$ channels. For the $\DD$ systems, in addition to the $T_{cc}$ state as the S-wave isospin singlet, its P-wave partner with $J^P=0^-$ also emerges as a near-threshold resonance. P-wave resonance poles are obtained in the $1^{-}$ and $2^-$ channels but are distant from the thresholds.

Concerning $I=1$ channels, besides a virtual state pole in the S-wave $1^{+-}$ channel, corresponding to the $Z_c(3900)$, no additional states are obtained. This is because the isospin factor $\tau\cdot\tau$ is only $\frac{1}{3}$ of the $I=0$ channels, rendering the potentials generally insufficient to generate bound states or P-wave resonances. For clarification, we do find resonance poles in $0^{-+}$, $1^{--}$ isovector $\DDbar$ channels and $1^{-}$, $2^{-}$ isovector $\DD$ channels, but they are very sensitive to the cutoff.  They transform into virtual states rather than bound states when adjusting the interaction strength, thereby  are not the kind of P-wave resonances we refer to in Fig.~\ref{fig:pole_taylor} (b). Their existence depends on the particular regularization, making them less credible. So we omit them in the final results.

To sum up, aside from the $\chi_{c1}(3872)$, $Z_c(3900)$, $G(3900)$, and $T_{cc}$ states, the S-wave $\DDbar$ state with $I^G (J^{PC})=0^- (1^{+-})$, P-wave $\DDbar$ state with $I^G(J^{PC})=0^+(0^{-+})$, and P-wave $\DD$ state with $I(J^P)=0(0^-)$ are more likely to be observed due to their proximity to the thresholds.

\emph{Conclusion and outlook.}---  The newly observed $G(3900)$ is interpreted as the P-wave $\DDbar$ resonance in a novel scenario. The existence of the P-wave resonance is firmly established on a unified meson-exchange model which well depicts the features of $\chi_{c1}(3872)$, $Z_c(3900)$ and $T_{cc}(3875)$ simultaneously. Compared with the S-wave state, the P-wave interaction dominantly arises from the well-known long-range pion exchange, ensuring robust conclusions when shifting between different models. The appearance of the P-wave resonance is also quite natural, particularly when the P-wave channels lack sufficient attraction, thus rendering them less sensitive to the potential shape compared with the S-wave resonance. This mechanism contributes to the dense population of P-wave resonances in both the $\DDbar$ and $\DD$ systems, which is  validated by our extensive calculations spanning all systems up to P-wave with various quantum numbers. 

Furthermore, there is promise in identifying P-wave resonances in other systems. For instance, the odd-parity $X_1(2900)$ observed in LHCb alongside  $X_0(2900)$~\cite{LHCb:2020bls,LHCb:2020pxc} may be plausibly interpreted as the P-wave $\bar{D}^\ast K^\ast$ resonance.  The $\psi(4220)$ state may potentially be interpreted as the P-wave $D_s^\ast {\bar D}_s^\ast$ resonance $Y(4220)$. Similarly, there may exist the P-wave $D_s^\ast {\bar D}_s/D_s {\bar D}_s^\ast$, $D_s {\bar D}_s$, $D^\ast {\bar D}^\ast$ and $D {\bar D}$ near-threshold resonances. One may also expect similar P-wave structures in the two bottom meson systems.

The $G(3900)$ being a P-wave resonance with the real part of the pole position below the $D\bar{D}^*$ threshold, its decaying to the $D\bar{D}^*$ final state is still allowable. Due to the imaginary part, the energy of the state becomes a distribution. There remains a significant probability that the state lies above the threshold, where the decay to the $D\bar{D}^*$ channel becomes kinematically allowed. Remarkably, for a resonance below the threshold, the line shape deviates severely from the Breit-Wigner form. In the future, more precise measurements of the $e^+e^-\to D\bar{D}^*$ process near the threshold could allow for an exclusive analysis using the $K$-matrix formalism to confirm the existence of the $G(3900)$. Additionally, searches for the $G(3900)$ can be conducted in hidden-charm final states, such as $J/\psi\eta$ and $\eta_c\omega$, with a relative P wave. The existence of the $G(3900)$ is a natural consequence of hadronic molecules like the $\chi_{c1}(3872)$, $T_{cc}(3875)$, and $Z_c(3900)$. Therefore, precise experimental measurements and refined theoretical calculations regarding the $G(3900)$ will provide valuable information and potentially direct constraints on the nature of the $\chi_{c1}(3872)$, $T_{cc}(3875)$, and $Z_c(3900)$.

Among the abundant predictions, the S-wave $\DDbar$ state with $I^G (J^{PC})=0^- (1^{+-})$, P-wave $\DDbar$ state with $I^G(J^{PC})=0^+(0^{-+})$, and P-wave $\DD$ state with $I(J^P)=0(0^-)$ are more likely to be observed due to their proximity to the thresholds. Unlike the $1^{--}$ $G(3900)$ state, their decay mode to  $D\bar{D}$ system is forbidden. Thus, these predictions could be searched in the hidden charmed channels, for example, the $1^{+-}$ state in $\eta_c\omega$, $J/\psi\eta$, $J/\psi\pi\pi$, the $0^{-+}$ state in $J/\psi\omega$, $\eta_c\pi\pi$ and $\chi_{c1}\pi\pi$, the $2^{-+}$ state in $J/\psi\omega$, $\chi_{c1}\pi\pi$, the $2^{--}$ state in $J/\psi\eta$, $\eta_c\omega$ etc.

\begin{acknowledgements}
The authors thank Yan-Ke Chen for helpful discussions. L.M is grateful to the helpful communications with Eric S. Swanson and Nils Hüsken. This project was supported by the National
Natural Science Foundation of China (11975033, 12147168 and 12070131001). This project was also funded by the Deutsche Forschungsgemeinschaft (DFG,
German Research Foundation, Project ID 196253076-TRR 110). J.Z.W. is also supported by the National Postdoctoral Program for Innovative Talent.

\end{acknowledgements}

\newcommand{\smref}{\cite{LHCb:2021auc,LHCb:2021vvq,Zhao:2014gqa,Du:2021zzh,Lin:2022wmj,Cheng:2022qcm,Albaladejo:2015lob,ParticleDataGroup:2024cfk,BESIII:2024ths,Julin:2017jcl,BESIII:2024ths,BESIII:2021yvc,CLEO:2008ojp,Dong:2017tpt,Belle:2017grj,Rapidis:1977cv,Schindler:1979rj,Osterheld:1986hw,BES:2001ckj,Ablikim:2006mb,Husken:2022yik}}

\bibliography{ref}


\clearpage

\setcounter{equation}{0}
\setcounter{figure}{0}
\setcounter{table}{0}
\setcounter{page}{1}
\makeatletter
\renewcommand{\theequation}{S\arabic{equation}}
\renewcommand{\thefigure}{S\arabic{figure}}
\renewcommand\thetable{SM-\Roman{table}}  

\begin{widetext}

\begin{center}
\textbf{\large Supplemental Materials: \\ \ttle}
\end{center}
\begin{mdframed}[hidealllines=true,innerleftmargin=0.1\textwidth,innerrightmargin=0.1\textwidth]
~~~~~This supplemental material provides additional details on \smabs.
\end{mdframed}

\end{widetext}

\section{The potentials}~\label{app:potentials}
\subsection{Sign problem of the $u$-channel potentials}~\label{app:qork}
In Fig.~\ref{fig:OBE} of the main text, the $u$-channel diagrams of the OBE are involved. However, a prevailing misconception exists in much of the literature regarding these $u$-channel OBE diagrams. We aim to address this misconception and illustrate its impact, revealing that while it does not introduce errors for the systems with even orbital angular momentum, it does induce a sign alteration in the partial wave potential for odd orbital angular momentum.

\begin{figure}[htbp]
  \centering
  \includegraphics[width=0.40\textwidth]{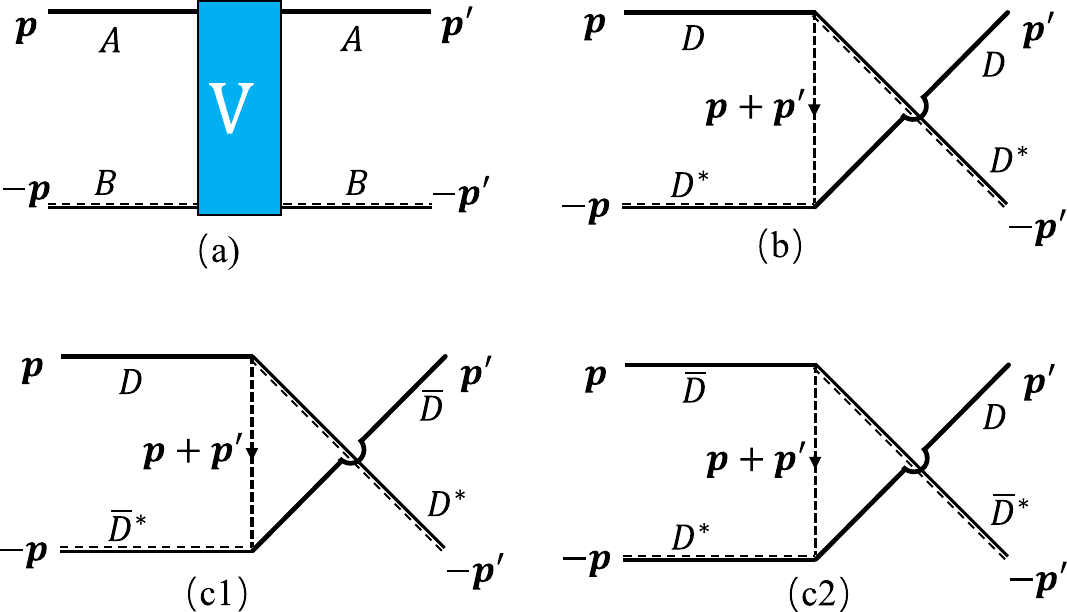} 
  \caption{\label{fig:allkernel} The momentum labeling for the $u$-channel diagrams. }
\end{figure}

In the general elastic scattering depicted for two distinguishable particles, denoted as $A$ and $B$, as illustrated in Fig.~\ref{fig:allkernel} (a), the nonlocal potential in the center-of-mass frame can be expressed as:
\begin{eqnarray}
        {V}(\bm{p}',\bm{p})\equiv\langle \bm p' |\hat{V}| \bm p \rangle \equiv \langle A(\bm{p}')B(-\bm{p}')|\hat{V}|A(\bm{p})B(-\bm{p})\rangle,\nonumber\\
\end{eqnarray}
Here, we adopt the momentum of the particle $A$ to label the two-particle states, denoted as $|\bm{p}\rangle$, which signifies $|A(\bm{p})B(-\bm{p})\rangle$. One can get the Lippmann-Schwinger equation in momentum space by sandwiching the operator equation $  \hat{T}=\hat{V}+\hat{V}\hat
{G}\hat{T}$ between initial and final two-body states and inserting complete basis between operators,
\begin{equation}
    T(\bm{p}',\bm{p};E)={V}(\bm{p}',\bm{p})+\int\frac{d\bm{p}''}{(2\pi)^{3}}{V}(\bm{p}',\bm{p}'')G(E,\bm{p
})T(\bm{p}'',\bm{p};E),~\label{eq:LSE}
\end{equation}
with
\begin{equation}
    G(E,\bm{p})=\frac{1}{E-m_{A}-m_{B}-\frac{p^{2}}{2m_{A}}-\frac{p^{2}}{2m_{B}}+i\epsilon}.
\end{equation}
It is worth noting that in Eq.~\eqref{eq:LSE}, all the three-momenta refer specifically to those of particle $A$.

Now we can specify $A$ and $B$ as $D$ and $D^*$, respectively taking OPE interaction as an example as shown in Fig~\ref{fig:allkernel} (b). Apparently, the momentum of the exchanged pion should be $\bm{k}=\bm p+ \bm p'$ rather than the conventional  $\bm q= \bm p'- \bm q$ used for the t-channel diagram. 

For the  $D^*\bar{D}/\bar{D}^*D$ system, one can construct the state with C-parity for the neutron channel,
\begin{equation}
    |D\bar{D}^{*}/\bar{D}D^{*},\{\beta,\bm{p}\}\rangle \equiv|D(\bm{p})\bar{D}^{*}(-\bm{p})+\beta\bar{D}(\bm{p})D^{*}(-\bm{p})\rangle.
\end{equation}
To ensure the fixed C-parity, the $\bar{D}(D^*)$ in the second component should has the same momentum as the $D(\bar{D}^*)$ in the fist component.  One can get
\begin{equation}
    \hat{C}|D\bar{D}^{*}/\bar{D}D^{*},\{\beta,\bm{p}\}\rangle=-\beta|D\bar{D}^{*}/\bar{D}D^{*},\{\beta,\bm{p}\}\rangle.
\end{equation}
with the convention
\begin{equation}
\hat{C}|D(\bm{p})\rangle=|\bar{D}(\bm{p})\rangle;\, \hat{C}|D^{*}(\bm{p})\rangle=-|\bar{D}^{*}(\bm{p})\rangle\label{eq:Cconvention}
\end{equation}
Taking OPE as an example, the Feynman diagrams involved are shown in Fig.~\ref{fig:allkernel} (c1) and (c2). Once again, the momentum of the exchanged pion should be $\bm{k}=\bm p+ \bm p'$.

As far as we know, in much of the literature discussing $u$-channel OBE diagrams, there is a common mistake regarding the momentum of the exchanged meson, where it is erroneously taken as $\bm{q} = \bm{p}' - \bm{p}$. Using $\bm{q}$ as $\bm{k}$ is equivalent to substituting $V(\bm{p}',\bm{p})$ with $V(\bm{p}',-\bm{p})$. Consequently, the potential term in the Schrödinger equation becomes:
\begin{eqnarray}
    &&\int d^{3}\bm{p}\ensuremath{V(\bm{p}',-\bm{p})\phi_{L}(\bm{p})}\nonumber\\
    &=&\int d^{3}\bm{p}\ensuremath{V(\bm{p}',\bm{p})\phi_{L}(-\bm{p})}\nonumber\\
    &=&(-1)^{L}\int d^{3}\bm{p}\ensuremath{V(\bm{p}',\bm{p})\phi_{L}(\bm{p})}
\end{eqnarray}
where $\phi_{L}(\bm{p})$ represents  wave function with orbital angular momentum $L$. One can see the mistake only affects states with odd $L$.  Fortunately, the majority of past literature has focused on S-wave and D-wave systems, rendering this mistake inconsequential for them. Recent studies concerning P-wave systems have acknowledged this issue and adopted the correct notations~\cite{Meng:2023bmz, Wang:2023ivd}.

\subsection{The OBE potentials and the G-parity rule\label{appendix:Veff}}
  \begin{figure*}
       \centering
       \includegraphics[width=1\textwidth]{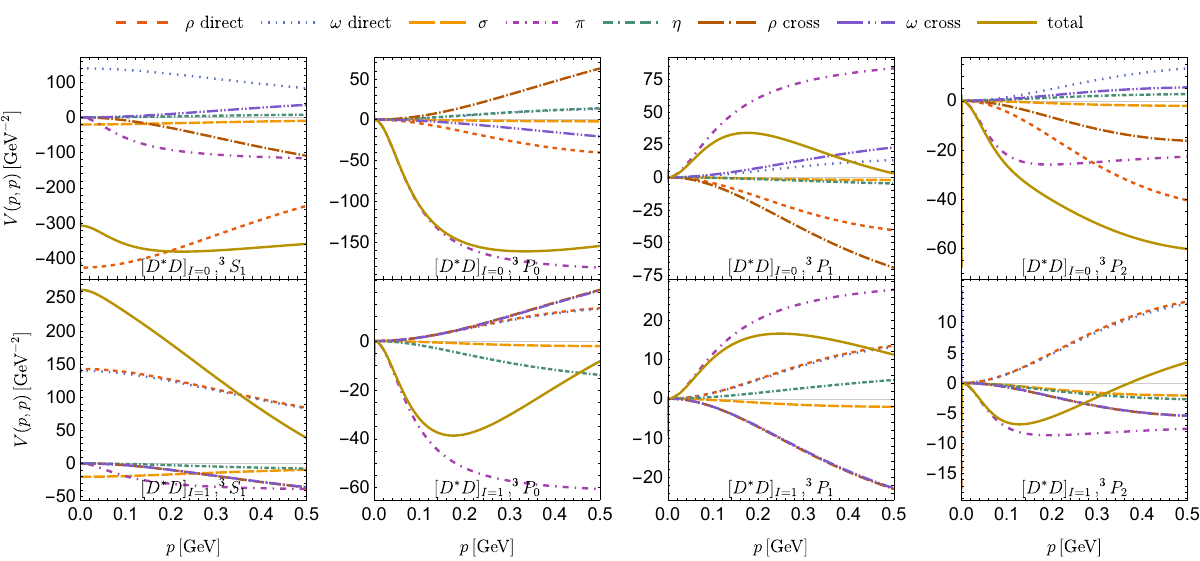}
       \caption{\label{fig:potentialDD} The OBE potentials without regulators of $\DD$ channels   are illustrated with various meson exchanges. Only $p=p'$ cases are depicted.}
   \end{figure*}

      \begin{figure*}
       \centering
       \includegraphics[width=1\textwidth]{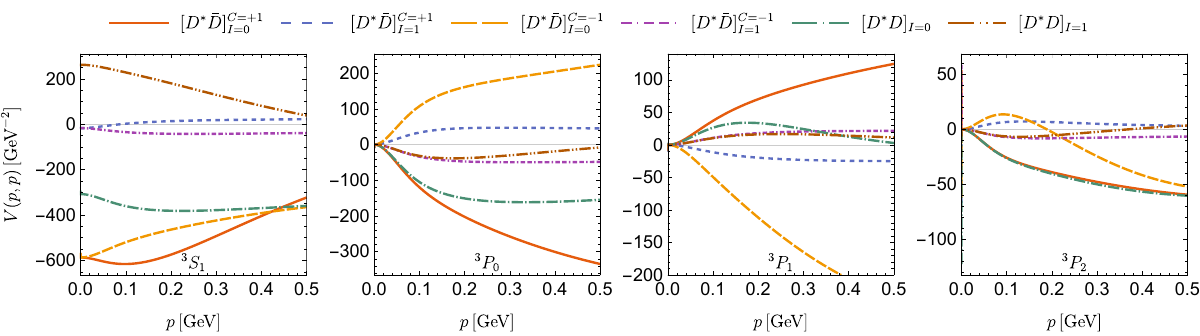}
       \caption{\label{fig:potentialall} The OBE potentials without regulators for all $\DD$ and $\DDbar$ systems up to P-wave with various quantum numbers. Only $p=p'$ cases are shown.}
   \end{figure*}

The effective potentials for the $DD^*$ system in momentum space are listed as follows,
\begin{eqnarray}
    &V_{\sigma}^{D}(\bm{p}',\bm{p})&=-\frac{g_{s}^{2}}{\bm{q}^{2}+m_{\sigma}^{2}},\nonumber\\
    &V_{\pi}^{C}(\bm{p}',\bm{p})&=-\frac{g^{2}}{2f_{\pi}^{2}}\frac{(\bm{\epsilon}\cdot \bm{k})(\bm{\epsilon}'\cdot \bm{k})}{\bm{k}^{2}-k_0^2+m_{\pi}^{2}}\tau\cdot\tau,\nonumber\\
    &V_{\eta}^{C}(\bm{p}',\bm{p})&=-\frac{g^{2}}{6f_{\pi}^{2}}\frac{(\bm{\epsilon}\cdot \bm{k})(\bm{\epsilon}'\cdot \bm{k})}{\bm{k}^{2}-k_0^2+m_{\eta}^{2}}\mathbbm{1}\cdot\mathbbm{1},\nonumber\\
    &V_{\rho/\omega}^{D}(\bm{p}',\bm{p})&=\frac{\frac{1}{4}\beta^{2}g_{V}^{2}(\bm{\epsilon}\cdot\bm{\epsilon}')}{\bm{q}^{2}+m_{\rho/\omega}^{2}}\times\begin{cases}
        \tau\cdot\tau,\quad \text{for }\rho,\\
        \mathbbm{1}\cdot\mathbbm{1},\quad \text{for }\omega,
    \end{cases}\nonumber\\
    &V_{\rho/\omega}^{C}(\bm{p}',\bm{p})&=\frac{\lambda^{2}g_{V}^{2}}{\bm{k}^{2}-k_0^2+m_{\rho/\omega}^{2}}\{(\bm{k}\cdot\bm{\epsilon})(\bm{k}\cdot\bm{\epsilon}')\nonumber\\
    &&-\bm{k}^{2}(\bm{\epsilon}\cdot\bm{\epsilon}')\}\times\begin{cases}
        \tau\cdot\tau,\quad \text{for }\rho,\\
        \mathbbm{1}\cdot\mathbbm{1},\quad \text{for }\omega,
    \end{cases}
\end{eqnarray}
where $D$ and $C$ denotes the direct and cross diagrams, respectively. The isospin factors are 
\begin{eqnarray}
    \tau\cdot\tau=\begin{cases}
        1,\quad I=1,\text{ D},\\
        -3,\quad I=0,\text{ D},\\
        1,\quad I=1,\text{ C},\\
        3,\quad I=0,\text{ C},
    \end{cases}\quad
    \mathbbm{1}\cdot\mathbbm{1}=\begin{cases}
        1,\quad I=1,\text{ D},\\
        1,\quad I=0,\text{ D},\\
        1,\quad I=1,\text{ C},\\
        -1,\quad I=0,\text{ C}.
    \end{cases}
\end{eqnarray}
The results of the partial-wave expansion potential $V=(\bm{\epsilon}\cdot \bm{k})(\bm{\epsilon}'\cdot \bm{k})D(p',p,z)$ with $z=\bm p'\cdot \bm p/pp'$, are listed as follows,
\begin{eqnarray}
&V_{S}^{J=1}=&\frac{2\pi}{3}\int_{-1}^{1}D(p',p,z)(p^{2}+p'^{2}+2pp'z)dz,\nonumber\\
    &V_{P}^{J=0}=&2\pi\int_{-1}^{1}D(p',p,z)\{(p^{2}+p'^{2})z+pp'(1+z^{2})\}dz,\nonumber\\
    &V_{P}^{J=1}=&2\pi\int_{-1}^{1}D(p',p,z)\frac{1}{2}(z^{2}-1)pp'dz,\nonumber\\
    &V_{P}^{J=2}=&\frac{2\pi}{5}\int_{-1}^{1}D(p',p,z)\{2(p^{2}+p'^{2})z\nonumber\\
    &&+\frac{1}{2}pp'(1+7z^{2})\}dz.
\end{eqnarray}
In Fig.~\ref{fig:potentialDD}, we list the the OBE potentials of $\DD$ channels with various meson exchanges.

The effective potential for the $\DDbar$ system can be related to the $\DD$ potential up to a factor $G_{m}$ ($-G_m G_{MM}$) for the direct (cross) diagrams, with $G_m$ and $G_{MM}$ as the G-parities of the exchanged meson and the $\DDbar$ system, respectively. In Fig.~\ref{fig:potentialall}, the total potentials for all $\DD$ and $\DDbar$ up to P-wave with various quantum numbers are illustrated. One can read out the $\DDbar$ potential of the specific meson exchange from Fig.~\ref{fig:potentialDD} via G-parity rule. It is noteworthy that the G-parity rule for cross diagrams is different from that in direct diagrams. We will present the derivation of the G-parity rule as follows.

\begin{figure}[ht]
    \centering
    \includegraphics[width=0.48\textwidth]{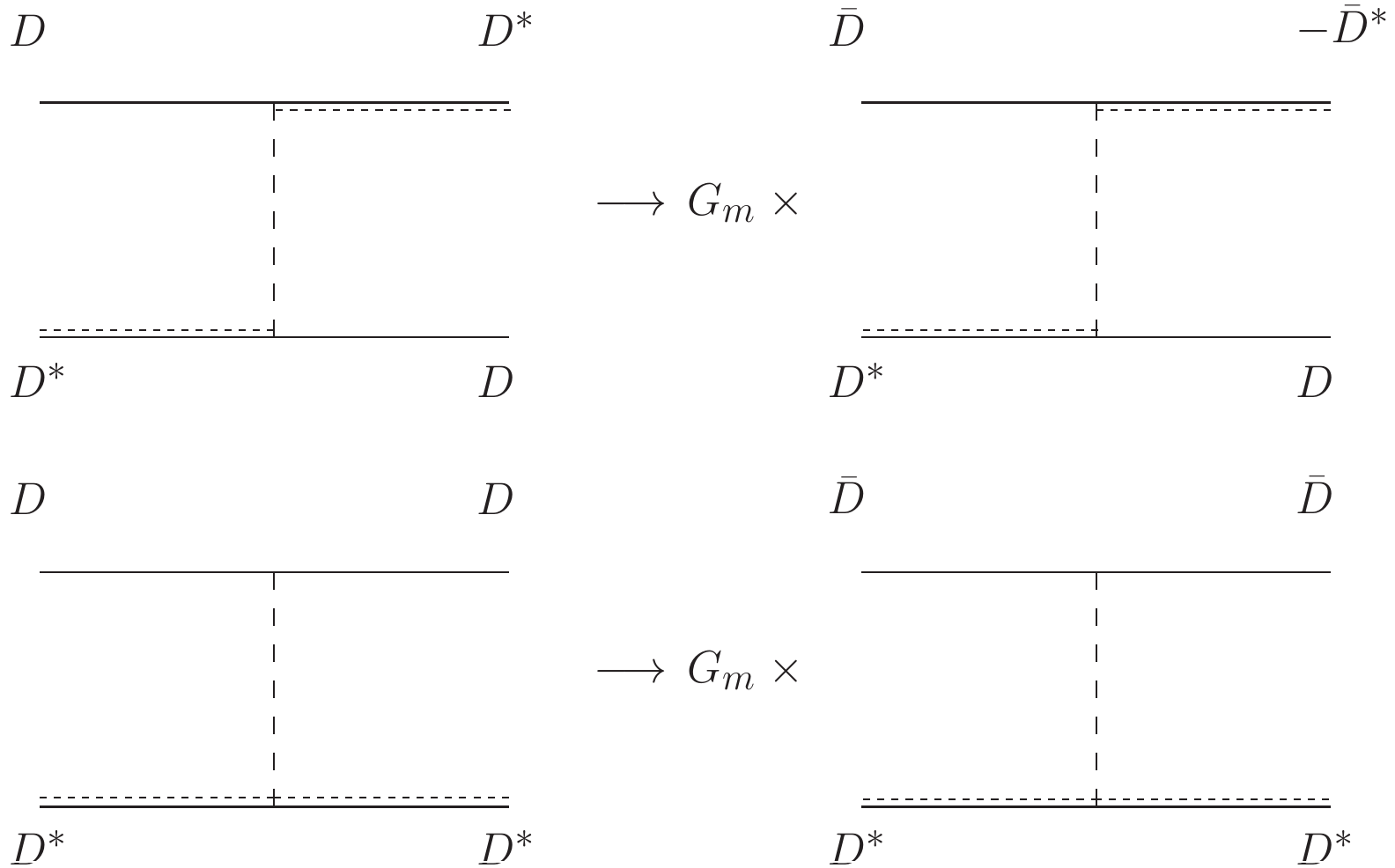}
    \caption{The G-parity transformation for cross and direct diagrams, respectively. The signs are determined by the G-parity of the exchanged meson and Eq.~(\ref{eq:Gconvention}).}
    \label{fig:Grule}
\end{figure}
With the C-parity convention in Eq.~(\ref{eq:Cconvention}) (the final result is irrelevant to the convention), the G-parity transformation reads
\begin{eqnarray}
    &&D=(-D^+,D^0)\xrightarrow{G}\bar{D}=(\bar{D}^0,D^-)\xrightarrow{G}-D,\nonumber\\
    &&D^*=(-D^{*+},D^{*0})\xrightarrow{G}-\bar{D}^*=-(\bar{D}^{*0},D^{*-})\xrightarrow{G}-D^*,\label{eq:Gconvention}\nonumber\\
\end{eqnarray}
where the charmed mesons are written in the form of isospin doublets. The G-parity eigenstates can be constructed,
\begin{eqnarray}
    |\DDbar,\,G=\pm\rangle=\frac{1}{\sqrt{2}}(|D\bar{D}^*\rangle\pm|\bar{D}D^*\rangle)\label{eq:Gstate}.
\end{eqnarray}
Since the exchanged mesons are eigenstates of the G-parity, we can apply the G-parity transformation to one of the vertex in $DD^*\rightarrow DD^*$, as shown in Fig.~\ref{fig:Grule}. The overall factor $G_m$ arises from the G-parity of the exchanged meson. Then we derive 
\begin{eqnarray}
    &&V^C_{\bar{D}D^*\rightarrow D\bar{D}^*}=(-G_m)V^C_{DD^*\rightarrow DD^*},\nonumber\\
    &&V^D_{\bar{D}D^*\rightarrow \bar{D}D^*}=V^D_{DD^*\rightarrow DD^*}.\label{eq:GtransformV}
\end{eqnarray}
Combining Eq.~(\ref{eq:Gstate}) and Eq.~(\ref{eq:GtransformV}), we derive
\begin{eqnarray}
    V_{\DDbar,\,G_{MM}}=G_m V^D_{DD^*\rightarrow DD^*}-G_m G_{MM} V^C_{DD^*\rightarrow DD^*}.\nonumber\\
\end{eqnarray}

\section{Identify the $G(3900)$ pole in the $K$-matrix formalism}
Very recently, a $K$-matrix formalism was used to analyze the $e^+e^-\to D\bar{D}$ data, in conjunction with the $e^+e^-\to D^*\bar{D}$, $e^+e^-\to D^*\bar{D}^*$, and $e^+e^-\to \text{everything}$ data~\cite{Husken:2024hmi}. This framework closely resembles that used in the bottom sector~\cite{Husken:2022yik}. The study concluded that no additional bare pole near 3900 MeV is needed to explain the data. However, the $K$-matrix formalism adopted in Ref.~\cite{Husken:2024hmi} differs from the standard one presented in the Review of Particle Physics. We have identified at least three defects in the analyses presented in Refs.~\cite{Husken:2022yik, Husken:2024hmi}.

The first defect is that the analyticity of the S-matrix is not maintained in Refs.~\cite{Husken:2022yik, Husken:2024hmi}. The center of mass momentum \( k \) for a two-body channel with masses \( m_1 \) and \( m_2 \) is given by:
\[
k(s)=\frac{\sqrt{[s-(m_1-m_2)^2][s-(m_1+m_2)^2]}}{2\sqrt{s}}.
\]
In the $K$-matrix formalism, the dependence on \( s \) of the amplitude is connected to the momenta \( k_i \) of different channels \( i \). In principle, one can analytically continue this definition to the complex plane of \( s \), considering that the analyticity of the S-matrix is a consequence of causality. However, in Ref.~\cite{Husken:2024hmi}, when calculating the cross sections, the momentum below the threshold is set to zero, namely introducing a discontinuous Heaviside step function. We believe this treatment violates the analyticity of the S-matrix and thus prevents the correct extraction of pole information, especially for poles that are somewhat far from the physical region.

The second defect is the subtraction-dependence in deriving Chew-Mandelstam function is not removed in the P-wave case. In the setting of Ref.~\cite{Husken:2024hmi}, for the P-wave channel, the subtracting constant in the Chew-Mandelstam function cannot be absorbed by the $K^{-1}$, leading to regularization dependence. The Chew-Mandelstam function is derived via once-subtracted dispersion relations. In principle, the subtraction point can be chosen freely because the constant at the subtraction point can be absorbed by \(K^{-1}\). However, in Refs.~\cite{Husken:2022yik,Husken:2024hmi}, for the P-wave (and higher partial wave) channels, the subtracting constant cannot be absorbed by \(K\) due to the \((k/\beta)^l\) factor in \(K\) introduced to satisfy the threshold behavior. Ensuring that the subtraction constant can be absorbed by \(K\) is equivalent to guarantee that physical observables are independent of regularization, which however is not ensured in Ref.~\cite{Husken:2022yik,Husken:2024hmi}.

Thirdly, in Refs.~\cite{Husken:2022yik, Husken:2024hmi}, for poles on the unphysical Riemann sheets, only the region above the threshold is considered. However, poles below but close to the threshold also have physical significance.

\begin{figure*}
    \centering
    \includegraphics[width=0.8\textwidth]{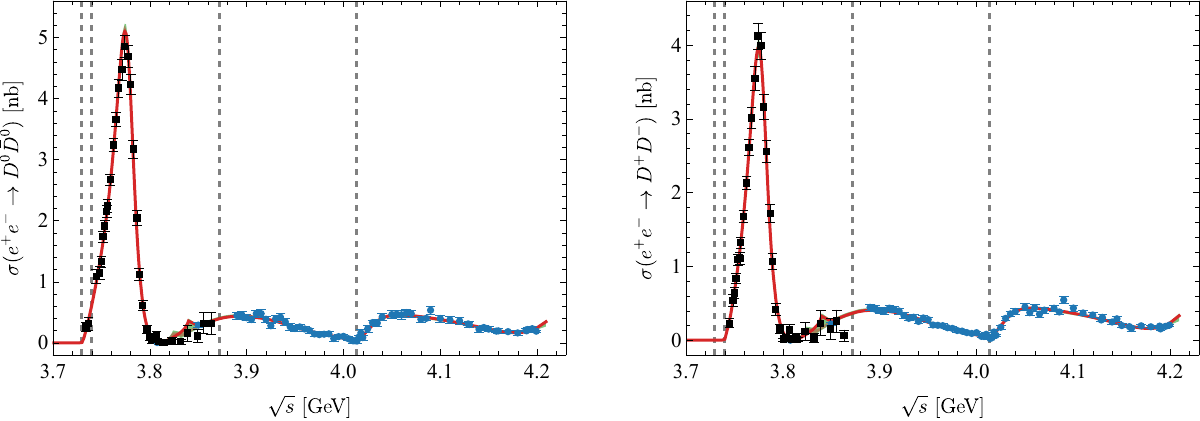}
        \includegraphics[width=0.8\textwidth]{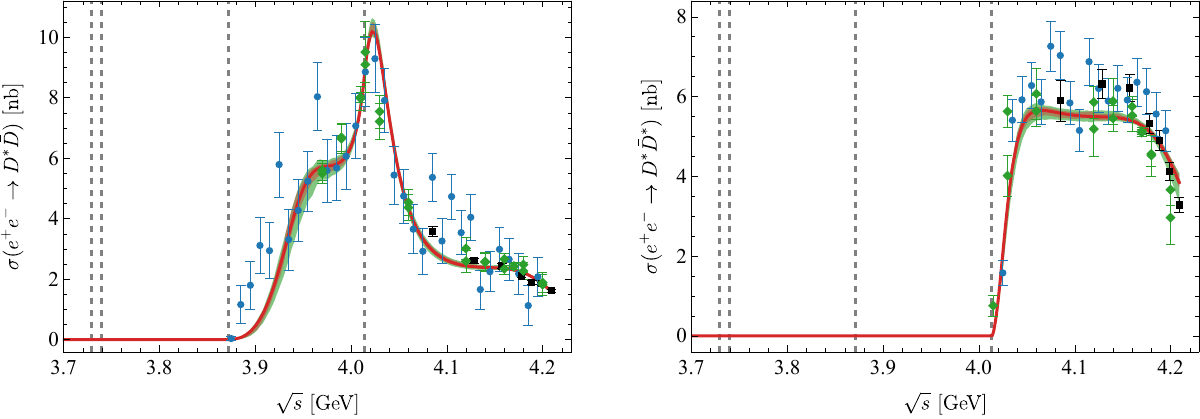}
           \includegraphics[width=0.8\textwidth]{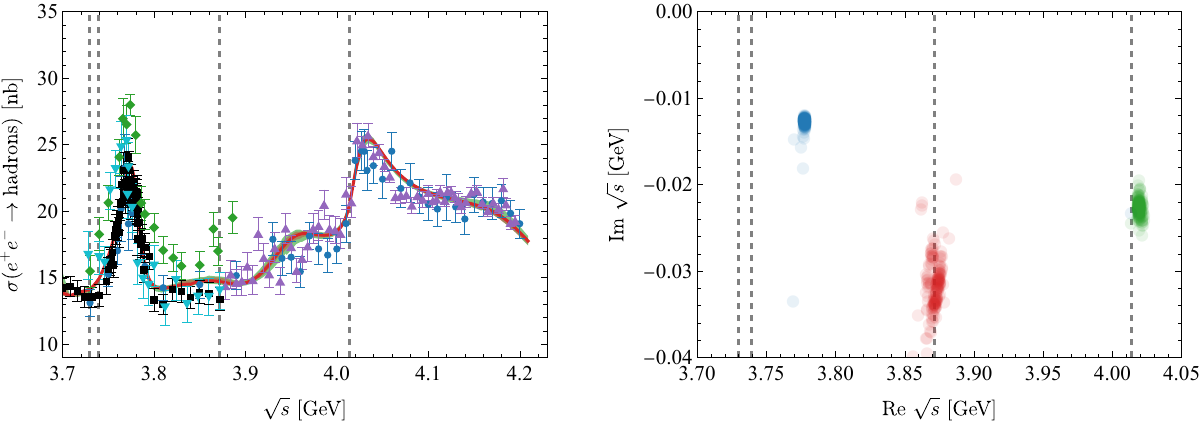}
    \caption{Our fit results in a modified scheme. The first five subfigures are the fit results for the line shapes, where  the red region and green region indicate the 68\% and 90\% confidence levels, respectively. The $D\bar{D}$ data is from BSEIII~\cite{Julin:2017jcl,BESIII:2024ths}. The $D^*\bar{D}$ and $D^*\bar{D}^*$ data is from BESIII~\cite{BESIII:2021yvc}, CLEO-c~\cite{CLEO:2008ojp,Dong:2017tpt}, and Belle~\cite{Belle:2017grj}. The inclusive data is from SPEAR~\cite{Rapidis:1977cv,Schindler:1979rj,Osterheld:1986hw}, BES ~\cite{BES:2001ckj}, and BESII~\cite{Ablikim:2006mb}.      
    The pole positions and uncertainties of \(\psi(3770)\), \(\psi(4040)\), and $G(3900)$ are depicted in blue, green, and red, respectively, in the subfigure of the third row and second column. The pole of $G(3900)$ is located in the \((-,-,-,+,-)\) sheet, labeled by the signs of the imaginary part of the momenta in the \(D^0\bar{D}^0\), \(D^+D^-\), \(\bar{D}D^*\), \(D^*\bar{D}^*\), and \(J/\psi(2\pi)\) channels in order.}
    \label{fig:myfit}
\end{figure*}

Given the possible defects mentioned above, we refit the data using a modified scheme. The main differences are: 
\begin{enumerate}
    \item We use consistent momentum in amplitudes to fit the cross sections and search for poles, namely, performing the analytical continuation to the complex plane;
        \item We change the regulator in $K$-matrix, $ e^{-k^{2}/\beta^{2}}\to[1+ k^2/\beta^2]^{-1} $
        to  avoid unphysical amplification for very negative $k^2$;
    \item We introduce five subtraction constants for \(D^0\bar{D}^0\), \(D^+D^-\), \(D^*\bar{D}\), \(D^*\bar{D}^*\), and a dummy channel as fitting parameters;
    \item In addition to the poles above the threshold, we also search for poles below but close to the threshold in our calculations.
\end{enumerate}
In Ref.~\cite{Husken:2024hmi}, five different models are used to explore various settings for isospin symmetry, node points in the wave function, and the dummy channel. It has been demonstrated that these different choices do not qualitatively affect the pole structures. Therefore, in our calculation, we use a single specific option: we choose the \(J/\psi (2\pi)\) dummy channel, assume the isospin symmetry for coupling constants, and neglect the node in the wave function. Considering extra five subtraction parameters, we perform a fit with 29 parameters. We vary \(\beta\) with values of 0.5, 0.7, and 1.0 GeV. The results are qualitatively consistent, with the fit for \(\beta = 0.7\) GeV yielding the smallest \(\chi^2\). Therefore, we present the results for \(\beta = 0.7\) GeV in Fig.~\ref{fig:myfit} and Table~\ref{tab:my_fit}. 

We find that the \(\chi^2/\text{dof}\) of our fit is actually smaller than those of four out of five models in Ref.~\cite{Husken:2024hmi}. Our results indicate the presence of a pole near the \(\bar{D}D^*\) thresholds. The width of this pole is considerably narrower than the result obtained by BESIII, suggesting that the Breit-Wigner fit used by BESIII may not be suitable for describing a near-threshold state. Interestingly, the pole position determined in our fit agrees well with that reported in Ref.~\cite{Du:2016qcr}, where a Lippmann-Schwinger equation formalism was used to analyze data from Belle. Both the Lippmann-Schwinger equation formalism and our $K$-matrix formalism comprehensively incorporate unitarity, analyticity, and threshold effects. Therefore, our refined results, in conjunction with those from Ref.~\cite{Du:2016qcr}, strongly support the existence of the $G(3900)$ pole.

\begin{table*}[htp]
    \centering
        \caption{The pole positions and $\chi^2$ in our fit (in units of MeV). The Ex. represent the experimental data from Refs.~\cite{ParticleDataGroup:2024cfk,BESIII:2024ths}.}
    \label{tab:my_fit}
	\begin{tabular*}{\hsize}{@{}@{\extracolsep{\fill}}ccccc@{}}
		\hline \hline 
 & $\psi(3770)$ & $\psi(4040)$ & $G(3900)$ & $\text{\ensuremath{\chi^{2}/\text{dof}}}$\tabularnewline
\hline 
Our fit & $3778.0(3)-i12.3(3)$ & $4019.5(5)-i22.9(11)$ & $3869.2(67)-i29.0(52)$ & 2.07\tabularnewline
Ex. & $3773.7(7)-i13.6(1)$ & $4040(4)-i42(6)$ & $3872.5(142)(30)-i89.9(70)(25)$ & \tabularnewline
\hline \hline 
\end{tabular*}
\end{table*}

\section{Systemic Uncertainties}~\label{app:sys_uncertainty}

\subsection{Uncertainties from the coupling constants}
In the main text, the coupling constants are determined using several models. It should be noted that these parameters were fixed before the observation of $G(3900)$ and $T_{cc}(3875)$. Therefore, the results for $G(3900)$ and $T_{cc}(3875)$ should be regarded as predictions to some extent.

\begin{figure}[htp]  \includegraphics[width=0.48\textwidth]{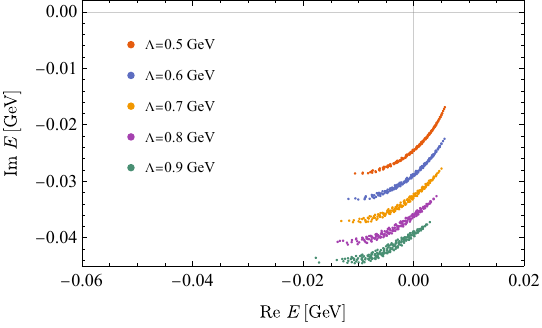}
\caption{$G(3900)$ pole positions determined by the coupling constants fixed by the $\chi_{c1}(3872)$, $Z_c(3900)$ and $T_{cc}(3875)$ pole positions in Eq.~\eqref{eq:XTZpole}.}\label{fig:poleII}
\end{figure}

To estimate the uncertainties of these coupling constants, we can redetermine these parameters using an alternative method. There are two independent couplings for the vector-meson exchange and one for the scalar-meson exchange. These three parameters can be fixed by the pole positions of $\chi_{c1}(3872)$, $Z_c(3900)$, and $T_{cc}(3875)$. We consider their positions within the following ranges:
\begin{eqnarray}
    \chi_{c1}(3872):~&& (-4,0)^{B}\text{ MeV},\nonumber\\
    T_{cc}(3875):~&& (-4,0)^{B}\text{ MeV},\nonumber\\
    Z_{c}(3900):~&&(-35,-15)^{V}\text{ MeV},~\label{eq:XTZpole}
\end{eqnarray}
where $B$ and $V$ represent the bound state and virtual state, respectively. The pole position range of $Z_c(3900)$ is motivated by Ref.~\cite{Albaladejo:2015lob}. We vary the cutoff from 0.5 GeV to 0.9 GeV in increments of 0.1 GeV. For each cutoff, we randomly select 200 sets of pole positions within the above ranges and then fix the three coupling constants. We find the obtained coupling constants vary within 30\% of those in the main text. The $G(3900)$ poles determined using these coupling constants are shown in Fig.~\ref{fig:poleII}. It is evident that a pole always exists close to the $D^*\bar{D}$ threshold. Thus, the conclusions in the main text remain unchanged when using this new method to determine the coupling constants.

\color{black}

\subsection{Regulator dependence}~\label{app:localR}

We also estimate the uncertainty of our results arising from the different regulators. For example, we choose the following regulators for the direct and cross diagrams, respectively,
\begin{eqnarray}
  &&  V^D(\bm{q})\rightarrow V^D(\bm{q})\left(\frac{\Lambda^2-m^2}{\Lambda^2+\bm{q}^2}\right )^2,\\\nonumber
    &&    V^C(\bm{k})\rightarrow V^C(\bm{k})\left(\frac{\Lambda^2-m^2}{\Lambda^2+\bm{k}^2}\right )^2, \label{eq:localR}
\end{eqnarray}
where the potentials from the direct and cross diagrams are the functions of $\bm q$ and $\bm k$, respectively. $m$ is the mass of the transferred meson. The cutoff $\Lambda=1.25$ and $1.35$ GeV are adjusted to get the loosely bound state $\chi_{c1}(3872)$. The pole positions for all other channels are presented in Table~\ref{tab:localR}. One can see as long as the cutoff is set to generate a loosely bound $\chi_{c1}(3872)$ state, a corresponding P-wave resonance emerges in the $1^{--}$ channel as the $G(3900)$, while the poles of $T_{cc}$ and $Z_c(3900)$ remain qualitatively unchanged. Our predictions are robust under various regularization schemes.

\begin{table*}[htpb]
	\centering
	\caption{\label{tab:localR}
		The poles in all channels of $D\bar{D}^*$ and $DD^*$, up to the orbital angular momentum $L=1$ with the regularization in Eq.~\eqref{eq:localR} (in units of MeV). The $B$ and $V$ superscripts denote the bound state and the virtual state, respectively. Otherwise the pole refers to a resonance. }	
	\begin{tabular*}{\hsize}{@{}@{\extracolsep{\fill}}cccccccc@{}}

		\hline \hline 
		&  & \multicolumn{2}{c}{$D\bar{D}^{*}\,,C=+$} & \multicolumn{2}{c}{$D\bar{D}^{*}\,,C=-$} & \multicolumn{2}{c}{$DD^{*}$}\tabularnewline
		&  & $I=0$ & $I=1$ & $I=0$ & $I=1$ & $I=0$ & $I=1$\tabularnewline
		\hline 
		\multirow{4}{*}{$\Lambda=1.25$ GeV} & $1^{+}({}^{3}S_{1})$ & $-0.40^B$, $\chi_{c1}(3872)$ & - & $-25.0^V$ & $-39.6^V$, $Z_{c}(3900)$ & $-0.79^B$, $T_{cc}(3875)$ & -\tabularnewline
		& $0^{-}({}^{3}P_{0})$ & $3.3-17.2i$ & - & - & - & $-11.2-16.7i$ & - \tabularnewline
		& $1^{-}({}^{3}P_{1})$ & - & - & $4.4-39.9i$, $G(3900)$ & - & $-96.6-87.3i$ & -\tabularnewline
		& $2^{-}({}^{3}P_{2})$ & $-71.2-63.5i$ & - &  $-31.0-96.5i$ & - & $-61.3-53.6i$ & -\tabularnewline
		\hline 
		\multirow{4}{*}{$\Lambda=1.35$ GeV} & $1^{+}({}^{3}S_{1})$ & $-2.8^{B}$, $\chi_{c1}(3872)$ & - & $-13.1^{V}$ & $-38.5^{V}$, $Z_{c}(3900)$ & $-8.8^{B}$, $T_{cc}(3875)$ & -\tabularnewline
		& $0^{-}({}^{3}P_{0})$ & $6.6-11.6i$ & - & - & - & $-10.2-18.0i$ & -\tabularnewline
		& $1^{-}({}^{3}P_{1})$ & -   & - & $10.2-33.7i$, $G(3900)$ & - & $-92.9-97.7i$ & -\tabularnewline
		& $2^{-}({}^{3}P_{2})$  &  $-68.0 - 75.4 i$ & - & $-23.3-97.2i$ & - & $-58.4-59.6$i & -\tabularnewline
		\hline 
		\hline 
	\end{tabular*}
\end{table*}

\subsection{The coupled-channel effect and three-body effect }~\label{app:correction}
In the main text, we only consider the $D\bar{D}^*/\bar{D}D^*$ channel. By expanding the Lagrangians with the heavy quark spin symmetry shown in Eq. (2) of the main text, we can examine the pole behavior of $G(3900)$ in the coupled-channel framework involving the $D\bar{D}$, $D\bar{D}^*/\bar{D}D^*$ and $D^*\bar{D}^*$. The comparison between the poles of $G(3900)$ in the single-channel case and the coupled-channel case are summarized in Table \ref{couplechannel}. The results demonstrate that the near-threshold feature of the $G(3900)$ pole is preserved even after taking into account the coupled-channel effects..

For the $D\bar{D}^*(DD^*)$ scattering, the $D\bar{D}\pi(DD\pi)$ three-body effect has been proved to be important for understanding the width of $S$-wave $T_{cc}^+$ and $X(3872)$~\cite{Du:2021zzh,Lin:2022wmj,Cheng:2022qcm}. For example, the $T_{cc}$ state would be a very loosely bound $DD^*$ state if there were no three-body decay to $DD\pi$. The three-body effect gives rise to the width of the $T_{cc}$ at the order of tens of keV. The impact of the three-body $D\bar{D}\pi$ effect from the OPE on the pole of $P$-wave $G(3900)$ is presented in Table~\ref{threebody}.  The three-body effect shifts the pole position by several MeVs, which is small compared with the width of the $G(3900)$. Unlike the \( T_{cc} \) and \( \chi_{c1}(3872) \) cases, the three-body effect plays only a minor role and is thus negligible.

\begin{table}[ht]
    \centering
    \caption{The comparison between the poles of $G(3900)$ within the single-channel case and the coupled-channel calculation involving $D\bar{D}$, $D\bar{D}^*/\bar{D}D^*$ and $D^*\bar{D}^*$ (in units of MeV). \label{couplechannel} }
    \begin{tabular*}{\hsize}{@{}@{\extracolsep{\fill}}ccccc@{}}
        \hline \hline 
       $\Lambda$ (GeV)  & $0.5$ & $0.6$ & $0.7$   \\
        \hline 
     Single channel & $-4.0-27.3i$ & $2.0-27.3i$ & $7.0-24.4i$    \\
     \hline
    Coupled channel & $-0.1-25.0i$  & $6.3-21.4i$  & $9.7-12.6i$  \\ 
     \hline         \hline 
        \end{tabular*}
\end{table}

\begin{table}[ht]
    \centering
    \caption{The impact of the three-body $D\bar{D}\pi$ effect from the OPE of $D\bar{D}^*\to\bar{D}D^*$ on the pole of $G(3900)$ (in units of MeV). \label{threebody} }
    \begin{tabular*}{\hsize}{@{}@{\extracolsep{\fill}}cccccccc@{}}
        \hline \hline 
       $\Lambda$ (GeV)  & 0.5 & 0.6 & 0.7   \\
        \hline 
     Without 3-body effect & $-4.0-27.3i$ & $2.0-27.3i$ & $7.0-24.4i$    \\
     \hline
    With 3-body effect & $-5.0-24.1i$  & $-1.2-24.3i$  & $6.3-21.7i$    \\ 
     \hline         \hline 
        \end{tabular*}
\end{table}

\subsection{The recoil correction and spin-orbit force}

\begin{figure*}[ht]
    \centering \includegraphics[width=0.99\textwidth]{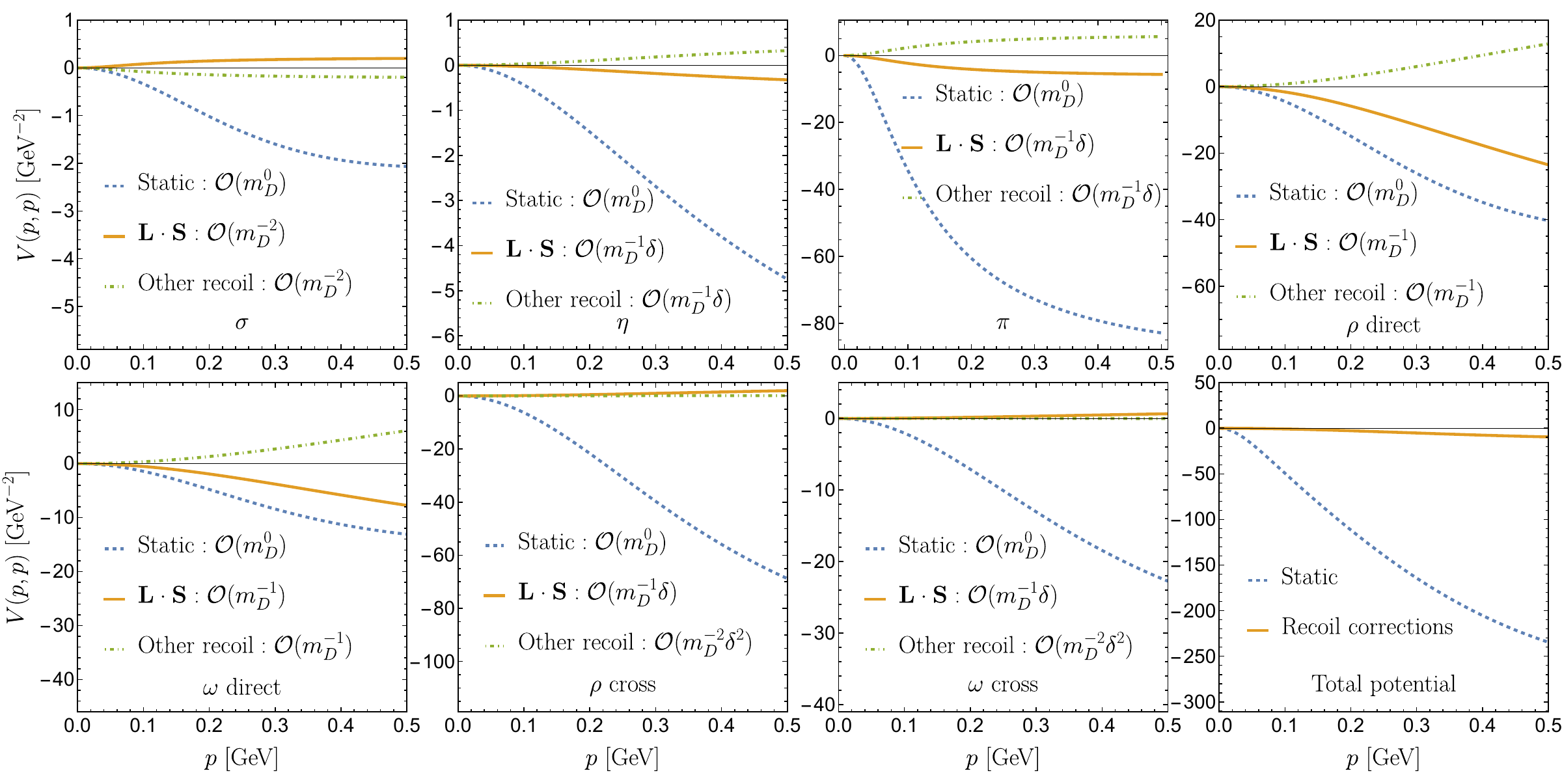}
\caption{The explicit recoil corrections for the $P$-wave effective potential of the $D\bar{D}^*$ associated with $G(3900)$ up to the order of $1/m_{D}^{2}$. Here, the $\delta=m_{D^*}-m_D$ is another small scale. Only $p=p'$ cases are depicted.}
    \label{fig:recoil}
\end{figure*}

In the main text, the effective potential is derived in the heavy meson limit, (see Appendix~\ref{app:potentials}), so the recoil corrections, which are suppressed by $1/m_{D^{(*)}}$, are neglected for consistency. The resulting effective potential in momentum space solely depends on the exchanged momentum $\bm q$ or $\bm k$. 
Here, in order to estimate the impact of this approximation, we investigate the recoil correction contribution \cite{Zhao:2014gqa} including the spin-orbit force for the $P$-wave effective potential of $D\bar{D}^*$ associated with $G(3900)$.

The polarization vector
of the vector meson in its rest frame  is 
\begin{eqnarray}
    \epsilon_{\lambda}=(0,\bm  \epsilon_{\lambda}). \label{eq:polav}
\end{eqnarray}
In order to include all the momentum-related terms, the
polarization vector in the laboratory frame is introduced by performing a Lorentz Boost to Eq. (\ref{eq:polav}), i.e., 
\begin{eqnarray}
  \epsilon_{\lambda}^{\mathrm{lab}}=\left(\frac{\bm p \cdot \bm \epsilon_{\lambda}}{m},\bm \epsilon_{\lambda}+\frac{(\bm p \cdot \bm \epsilon_{\lambda})\bm p}{(p_0+m)m}\right).  \label{eq:polavlab}
\end{eqnarray}
From Eq. (\ref{eq:polavlab}), we only keep the recoil corrections up to the order of $1/m_{D}^2$, whose explicit expressions for the $DD^*$ system are summarized as follow 
\begin{eqnarray}
    &&V_{\sigma}^{D(recoil)}(\bm{p}',\bm{p})=-\frac{1}{2m_{D^*}^2}\frac{g_{s}^{2}}{\bm{q}^{2}+m_{\sigma}^{2}}((\bm{\epsilon}\cdot \bm{q})(\bm{\epsilon}'\cdot \bm{q})-i\bm S \cdot\bm L),\nonumber\\
    &&V_{\pi/\eta}^{C(recoil)}(\bm{p}',\bm{p})=-\frac{g^{2}}{2f_{\pi}^{2}}\frac{1}{\bm{k}^{2}-k_0^2+m_{\pi/\eta}^{2}}
    \nonumber \\
    &&\quad \quad \quad \quad \times [-\frac{m_{D^*}^2-m_{D}^2}{2m_{D^*}^2}i\bm S \cdot\bm L \nonumber\\
    &&\quad \quad \quad \quad -\frac{(m_{D^*}-m_{D})^2(\bm{\epsilon}\cdot \bm{q})(\bm{\epsilon}'\cdot \bm{q})}{4m_{D^*}^2}-(\bm{\epsilon}\cdot \bm{k})(\bm{\epsilon}'\cdot \bm{k})  \nonumber\\
    && \quad \quad \quad \quad \times \frac{3m_{D^*}^2-2m_{D^*}m_{D}-m_{D}^2}{4m_{D^*}^2}]
    \times\begin{cases}
        \tau\cdot\tau,\quad \text{for }\pi,\\
        \mathbbm{1}\cdot\mathbbm{1},\quad \text{for }\eta,
    \end{cases}\nonumber\\    
    &&V_{\rho/\omega}^{D(recoil)}(\bm{p}',\bm{p})=\frac{1}{\bm{q}^{2}+m_{\rho/\omega}^{2}}
    \nonumber \\
    &&\quad \quad \quad \quad \times [(\frac{-\lambda \beta g_{V}^{2}}{2m_{D^*}}+\frac{\beta^{2}g_{V}^{2}}{8m_{D^*}^2})(\bm{\epsilon}\cdot \bm{q})(\bm{\epsilon}'\cdot \bm{q}) \nonumber \\
   &&\quad \quad \quad \quad +(\frac{\lambda \beta g_{V}^{2}(m_{D^*}+m_{D})}{2m_{D}m_{D^*}}-\frac{\beta^{2}g_{V}^{2}}{8m_{D^*}^2})i\bm S \cdot\bm L \nonumber \\
   &&\quad \quad \quad \quad +\frac{\beta^{2}g_{V}^{2}}{16m_{D^*}m_{D}}\bm{k}^{2}(\bm{\epsilon}'\cdot \bm{\epsilon})]\times\begin{cases}
        \tau\cdot\tau,\quad \text{for }\rho,\\
        \mathbbm{1}\cdot\mathbbm{1},\quad \text{for }\omega,
    \end{cases}\nonumber \\
    &&V_{\rho/\omega}^{C(recoil)}(\bm{p}',\bm{p})=\frac{1}{\bm{k}^{2}-k_0^2+m_{\rho/\omega}^{2}}
    \nonumber \\
    &&\quad \quad \quad \quad \times [\frac{\lambda^{2}g_{V}^{2}(m_{D^*}-m_{D})^2}{4m_{D^*}m_{D}}\bm{q}^{2}(\bm{\epsilon}'\cdot \bm{\epsilon}) \nonumber \\
    &&\quad \quad \quad \quad +\frac{\lambda^{2}g_{V}^{2}(2m_{D^*}+m_{D})(m_{D^*}-m_{D})^2}{4m_{D^*}^3}(\bm{q}\cdot\bm{\epsilon})(\bm{q}\cdot\bm{\epsilon}') \nonumber \\
    &&\quad \quad \quad \quad +\frac{\lambda^{2}g_{V}^{2}m_{D}(m_{D^*}^2-m_{D}^2)}{2m_{D^*}^3}i\bm S \cdot\bm L  \nonumber \\
    &&\quad \quad \quad \quad +\frac{\lambda^{2}g_{V}^{2}(m_{D^*}-m_{D})^2}{4m_{D^*}m_{D}}\times (\bm{k}\cdot\bm{\epsilon})(\bm{k}\cdot\bm{\epsilon}')   \nonumber \\
    &&\quad \quad \quad \quad +\frac{\lambda^{2}g_{V}^{2}(m_{D^*}-m_{D})^2(2m_{D^*}+m_{D})}{4m_{D^*}^3}
    \nonumber \\
    &&\quad \quad \quad \quad \times \bm{k}^2(\bm{\epsilon}\cdot \bm{\epsilon}')] \times\begin{cases}
        \tau\cdot\tau,\quad \text{for }\rho,\\
        \mathbbm{1}\cdot\mathbbm{1},\quad \text{for }\omega,
    \end{cases} \nonumber\\
 \end{eqnarray}  
where the spin operator and angular momentum operator are \(\bm{S} = -i(\bm{\epsilon} \times \bm{\epsilon}')\) and \(\bm{L} = (\bm{q} \times \bm{k})/2\), respectively. It can be seen that the spin-orbit force \(\bm{S} \cdot \bm{L}\) first appears at the order of \(1/m_{D}\). For the cross diagram, the recoil terms are always accompanied with a factor \(\delta\equiv m_{D^*} - m_{D}  \), which is a small scale and provides additional suppression compared with the  direct diagram.

In Fig. \ref{fig:recoil}, we present the partial-wave effective potential of the \(D\bar{D}^*\) associated with $G(3900)$ from the order \(\mathcal{O}(m_{D}^0)\) to \(\mathcal{O}(m_{D}^{-2})\). All higher-order recoil corrections can be divided into spin-orbit forces, tensor forces and central forces. Seen from Fig. \ref{fig:recoil}, the recoil effects, except for those of the direct \(\rho\) and \(\omega\) exchange potentials, are negligible compared with the leading static parts. For the direct \(\rho\) and \(\omega\) exchange potentials, the spin-orbit forces and the other recoil forces have different signs and almost cancel each other out, resulting in a small total correction. Therefore, we conclude that the effect of the recoil correction is minor.

   \section{Production and Line Shape}

\begin{figure}
    \centering
    \includegraphics[width=0.8\linewidth]{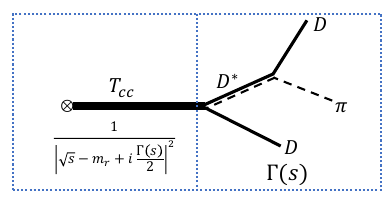}
    \caption{Feynman diagram related to the production line shape of $T_{cc}(3875)$. The diagram is divided into two parts separated in two boxes, the complete propagator of the $T_{cc}$ state with a production vertex labeled by ``$\otimes$", and its energy-dependence three-body decay width. }
    \label{fig:fey_prod}
\end{figure}

It is noteworthy that the \( T_{cc}(3875) \) was observed in the prompt production at the LHC~\cite{LHCb:2021auc,LHCb:2021vvq}. In the main text, several P-wave $D^*D$ resonances below the threshold are predicted. One might anticipate that prompt production is not selective regarding quantum numbers, and thus these P-wave resonances should also appear in the LHCb data. However, we will demonstrate that this is not the case using a simple calculation.

We will use the \(T_{cc}(3875)\) state and the P-wave state with the closest pole to the physical region, the isoscalar \(^3P_0\) \(DD^*\), as examples. The related Feynman diagram is presented in Fig.~\ref{fig:fey_prod}. The diagram is divided into two parts within two boxes: the complete propagator of the \(T_{cc}(3875)\) or P-wave state with production vertices labeled by “\(\otimes\),” and its energy-dependent three-body decay width. In prompt production, we assume the production vertices of S-wave and P-wave resonances are of the same order and neglect their differences in the following calculation. The energy-dependent decay width is calculated in the sequential formalism, where the couplings of the \(D^*\), \(D\), and dimeson resonances are determined by the residue of the \(DD^*\) scattering amplitude, specifically
\begin{eqnarray}
    g_{\text{eff}}=\langle k_R|\hat{V}|\phi\rangle,
\end{eqnarray}
or equivalently
\begin{eqnarray}
    g_{\text{eff}}^2=\lim_{E\to E_R} (E-E_R)T_{\text{on-shell}}(E),
\end{eqnarray}
where $g_{\text{eff}}$ is the effective vertex of the resonance and $DD^*$ (there is an extra momentum for P-wave states), $E_R=$ is the pole position, $k=\sqrt{2\mu E_R}$, the $|\phi\rangle$ is the resonance wave function obtained by the Schr\"odinger equation, and the $T_{\text{on-shell}}$ is the on-shell $T$ -matrix.

In the complete propagator of the dimeson resonance, \(m_r\) represents the real part of the related values in Table~\ref{tab:prediction}. To better meet the constraints of unitarity, the energy-dependent \(\Gamma\) evaluated via sequential decays is adopted.

\begin{figure*}
    \centering
  \includegraphics[width=0.99\textwidth]{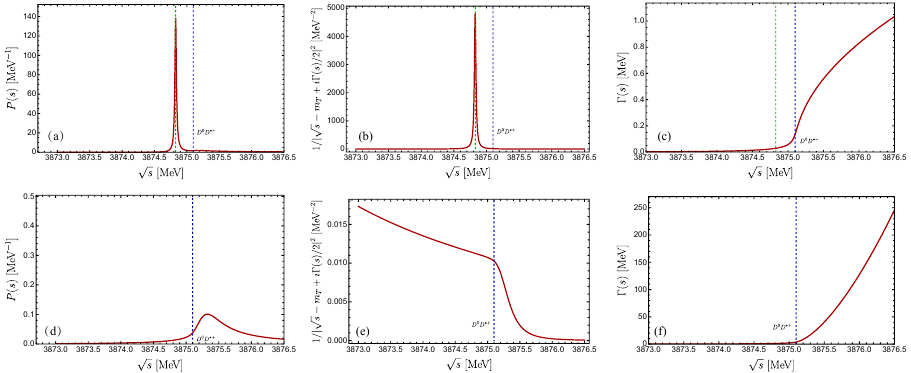}
    \caption{The production line shapes of $0(1^+)$ (upper panel) and $0(0^-)$ (lower panel) $D^*D$  dimeson states from the final state $D_0D_0\pi^+$. The $\Gamma(s)$ is energy dependent width and $P(s)=\Gamma(s)/|\sqrt{s}-m_T+i\Gamma(s)/2|^2$. The Green lines in (a),(b),(c) locate in the real part of the $T_{cc}^+$ pole.}~\label{fig:production_num}
\end{figure*}

\begin{figure}
    \centering
    \includegraphics[width=0.95\linewidth]{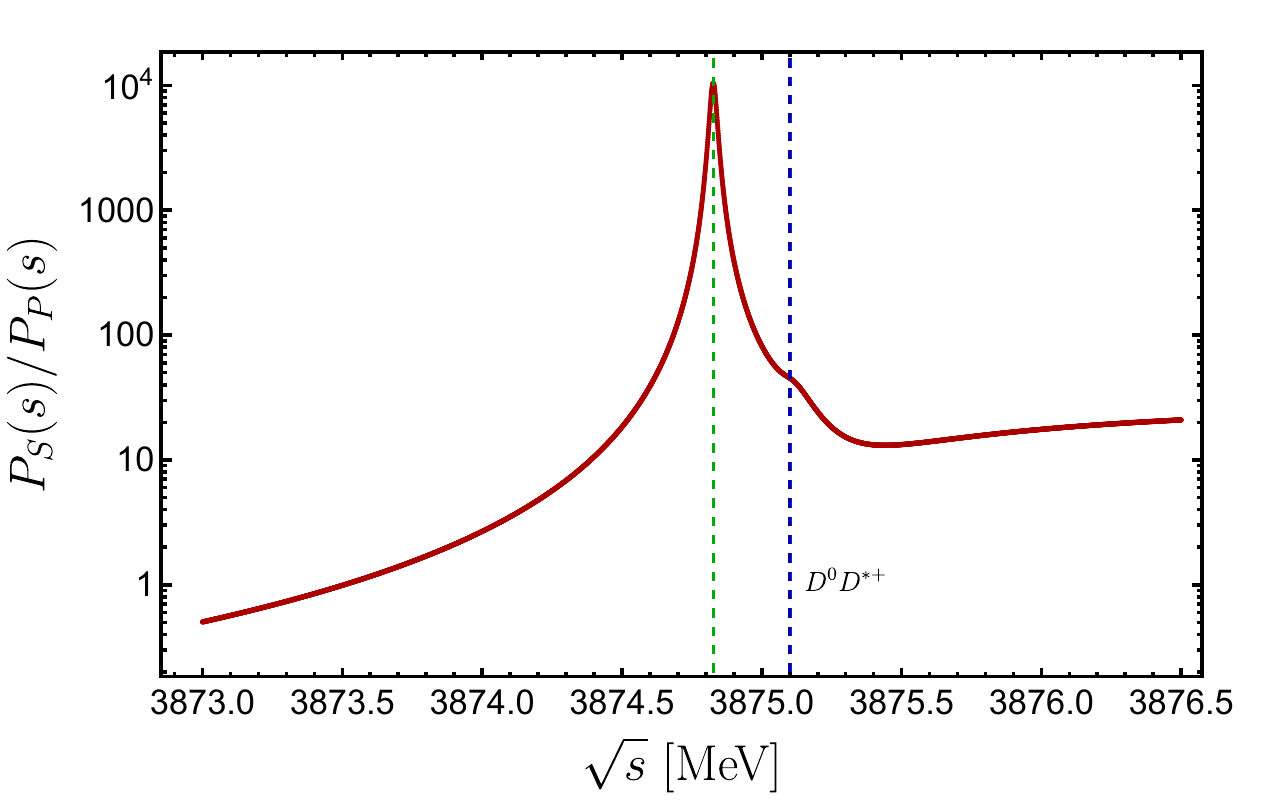}
    \caption{The production ratio $P_S(s)/P_P(s)$ between $0(1^+)$ and $0(0^-)$ state. 
	The Green line locates in the real part of the $T_{cc}^+$ pole.}
    \label{fig:production_ratio}
\end{figure}

Therefore, we define a quantity measuring the probability distribution function of the produced dimeson resonances,
\begin{equation}
    P_{S(P)}(s)={\Gamma(s) \over |\sqrt{s}-m_r+i \Gamma(s)/2|^2}.
\end{equation}
The subscript \(S\) or \(P\) represents the S-wave \(T_{cc}\) state or the P-wave resonance, respectively. If a constant \(\Gamma\) is used, the result is a Breit-Wigner distribution (also known as the Cauchy or Lorentz distribution). The integration of the Breit-Wigner distribution becomes a constant, independent of \(\Gamma\) and \(m_r\), which aligns with the expectation that prompt production is not selective regarding quantum numbers. However, when the energy dependence of \(\Gamma\) is considered, the situation changes, as shown in Figs.~\ref{fig:production_num} and \ref{fig:production_ratio}. Since the pole of the P-wave resonance is on the unphysical Riemann sheet and below the threshold, the magnitude of its complete propagator is significantly suppressed compared to the S-wave state. Therefore, the production probability density of the P-wave states is much smaller than that of the S-wave state by an order of magnitude, as shown in Fig.~\ref{fig:production_ratio}. We also evaluate the integrated production probability:
\[
\frac{\int P_S(s) \, d\sqrt{s}}{\int P_P(s) \, d\sqrt{s}} \approx 80.
\]

This indicates that the number of P-wave state events should be less than that of the S-wave by two orders. Therefore, considering the suppression in production, it is unlikely that the P-wave resonance can be identified in the current data.


\end{document}